\documentclass[aip, jmp,graphicx, amsmath, amssymb,twocolumn, reprint]{revtex4-1}
\usepackage{graphicx}
\usepackage{dcolumn}
\usepackage{bm}
\usepackage[mathlines]{lineno}
\draft 
\bibstyle{plain}
\bibdata{rhian}
\relax
\usepackage[breaklinks=true,colorlinks=true,linkcolor=blue,urlcolor=blue,citecolor=magenta]{hyperref}
\usepackage[caption=false]{subfig}
\usepackage{subeqnarray}
\begin{document}
\title{Demkov-Kunike Models with Decay}
\author{M. B. Kenmoe}
\affiliation{Mesoscopic and Multilayer Structures Laboratory, Faculty of Science, Department of Physics, University of Dschang, Cameroon}
\author{A. B. Tchapda}
\affiliation{Mesoscopic and Multilayer Structures Laboratory, Faculty of Science, Department of Physics, University of Dschang, Cameroon}
\author{L. C. Fai}
\affiliation{Mesoscopic and Multilayer Structures Laboratory, Faculty of Science, Department of Physics, University of Dschang, Cameroon}
\date{\today}
\begin{abstract}
Exact analytical solutions to the dissipative time-dependent Schr\"odinger equation are obtained for a decaying two-state system with decay rates $\Gamma_{1}$ and $\Gamma_{2}$ for levels with extremal spin projections. The system is coherently driven with a pulse whose detuning is made up of two parts: a time-dependent part (chirp) of hyperbolic-tangent shape and a static part with real and imaginary terms.  This gives us a wide range of possibilities to arbitrarily select the interaction terms. We considered two versions which led to decaying Demkov-Kunike (DK) models; the version in which the Rabi frequency (interaction) is a time-dependent hyperbolic-secant function (called decaying DK1 model) and the case when it is constant in time and never turns off (decaying DK2 model).  Our analytical solutions account for all possible initial moments instead of only $t_{0}=0$ or $t_{0}=-\infty$ as for non-decaying models and may be useful for experiments on level crossings.  Two complementary limits of the pulse detuning are considered and explored: the limit of fast (i)  and slow rise (ii). In the case (i), the coupling between level positions in the first DK model collapses while the second DK model reduces to a Rabi model (constant Hamiltonian), in the case (ii), both DK models reduce to the LZ model. In both cases (i) and (ii), analytical approximated solutions which conveniently approach the exact solutions are derived.
\end{abstract}
\maketitle
\section{Introduction}\label{Sec1}
The constantly increasing interest devoted to {\it non-resonant} two-level models has been triggered over the past decades by the discovery of their potential roles in modern and contemporary non-stationary quantum mechanics. To model and characterize some phenomena that occur at atomic and subatomic levels, several exactly solvable models have been proposed~\cite{lan, zen, stu, Majorana, Rosen, Rabi, Demkov, Bambini, Allen}. The most practical being the Landau-Zener model (LZ)~\cite{lan, zen, stu, Majorana} due to its simplicity and numerous applications~\cite{prigogine, sessoli1999.1, Averin1995, Averin1999, Farhi}.  However, few of these models are not reliable for experiments as they do not fully embed realistic situations and involve several drastic drawbacks.  In the LZ model for instance, the field is never turned off (the detuning strongly varies
linearly with time), diabatic energies (energies of uncoupled bare states) are infinite (not bounded) at large positive and negative times and the Rabi frequency (interaction) remains constant in time and never turns off.  Several attempts to reformulate this model and make it more realistic such that it  embeds surrounding effects have been published~\cite{kay1997, Kenmoe2013, pok2003, pok}. 

Demkov and Kunike have proposed two exactly solvable models which fulfill more physical requirements and stand as generalizations of the LZ model~\cite{Demkov}. These models assume a time-dependent detuning which has a static component and a hyperbolic-tangent chirp~\cite{Demkov, Garraway, Vitanov2007, Nakamura}. The first DK model denoted as DK1 is characterized by a bell-shaped pulse (see figure \ref{FIG1}) and a time-dependent interaction (Rabi frequency) of hyperbolic-secant shape while in the second DK model (DK2), the Rabi frequency is constant in time~\cite{Demkov, Garraway}.  Both models are convenient for finite-duration  pulses and have proven to be applicable in: the physics of ultracold trapped gases~\cite{Kohler}, superchemistry~\cite{Heinzen}, Bose-Einstein condensates~\cite{chin}, photo- and magneto- association for production of cold molecules~\cite{Sokhoyan}.

When  a negative imaginary term is added to the static part of the detuning, the total Hamiltonian becomes non-Hermitian (NH)\cite{Bender}. In this case, DK models are suitable  to describe excitations in the presence of fluorescence from the excited-state to other states outside the system (continuum), or in the presence of ionization from the excited-state induced by another coupling. Thus,  decaying DK models have attracted some remarkable attentions.   In Refs.[\onlinecite{Vitanov1997, Avishai2014}], exact asymptotic solutions at large positive times have been obtained for the generic case when only the excited-state of the system decays outside irreversibly. The work~[\onlinecite{Vitanov1997}] considered the first version (time-dependent hyperbolic-secant interaction) of DK models while~[\onlinecite{Avishai2014}] focused on the second version (constant interaction). In this paper, we consider the two versions of the DK model and mainly assume that not only one diabatic state (ground or excited)  decays outside irreversibly, but all. Exact analytical solutions that account for all possible initial time $t_{0}$ (turn-on time) instead of $t_{0}=-\infty$ (as considered by the authors in~[\onlinecite{Vitanov1997, Avishai2014}]) are obtained. Our solutions are valid for a wide range of atoms with different lifetimes in metastable states.  We have found that both models follow the same mathematical strategy and differ only by explicit values of  parameters  involved. Thus, we construct a theory which simultaneously copes with both models.

The structure of the paper is as follows. In sections \ref{Sec2} and \ref{Sec3}, we present our models and elaborate the theory presenting our strategy. In section \ref{Sec4.0}, the strategy is shown in action for the first and second decaying DK models. Section \ref{Sec6} considers two complementary limiting cases (that of fast and slow rise) while section \ref{Sec8} summarizes our main achievements. Extra appendices are provided: appendix \ref{App1} and  \ref{App2} sketch mathematical instruments and functions used for analytical calculations. Appendix \ref{App3} presents probability amplitudes obtained in the slow rise limit.  

\section{Models and Eigen-energies}\label{Sec2}
\subsection{Models}\label{Sec2.1}
\begin{figure}[!h]
\vspace{0.0cm}
 \begin{center}
					\includegraphics[width=7.cm, height=60mm]{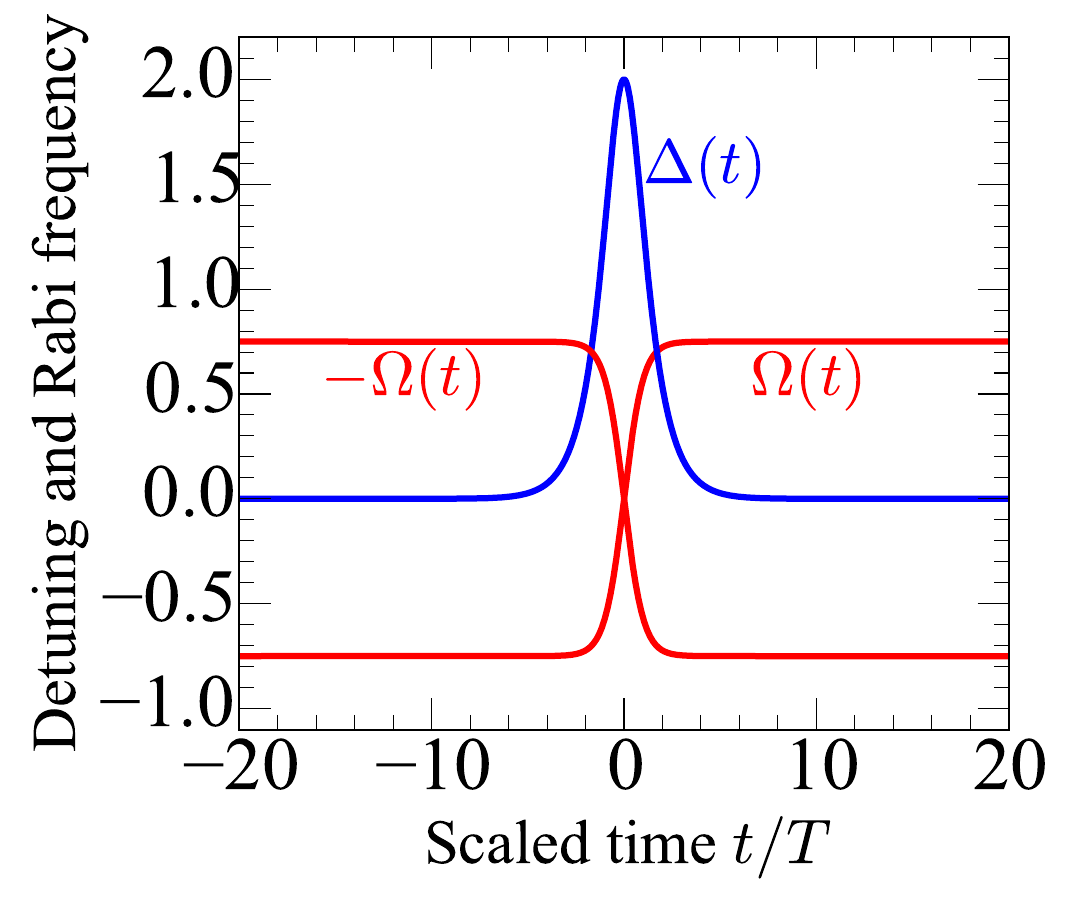} 
\end{center}
\vspace{-0.6cm}
\caption{Time-evolution of the detuning $\Omega(t)$ and the time-dependent Rabi frequency $\Delta(t)$ for DK1 plotted with $\Omega_{0}=0.75/T$, $D=0.0$ and $\Delta_{0}=2/T$. Two interesting driving regimes may be identified. In the circumstances, when $|t/T|\gg1$ (fast rise regime), the detuning saturates around  a single value, when $|t/T|\ll1$ (slow rise regime), it is nearly linear. 
} 
\label{FIG1}
\end{figure}

Consider a decaying two-state system (atom or molecule) with decay rates (inverse life times) $\Gamma_{1}$ and $\Gamma_{2}$ for diabatic states $|1\rangle$  and $|2\rangle$. They are  associated with wave functions $\psi_{1}(t,t_{0})$ and $\psi_{2}(t,t_{0})$ that are coherently driven from an initial time $t_{0}\le0$ to an arbitrary time $t\ge t_{0}$ by an external pulse of detuning $\Omega(t)$ (difference between the system's transition frequency and that of the  external field) and Rabi frequency $\Delta(t)$ which quantifies the field-induced coupling between $\psi_{1}(t,t_{0})$ and $\psi_{2}(t,t_{0})$. In the rotating-wave approximation, the corresponding probability amplitudes $C_{1,2}(t,t_{0})$ obey the dissipative time-dependent Schr\"odinger equation (in the units $\hbar=1$)
\begin{eqnarray}\label{equ1}
 i\frac{d\mathbf{C}(t)}{dt}
=\mathbf{H}(t)\mathbf{C}(t). 
\end{eqnarray}
Here, $\mathbf{C}(t)=[C_{1}(t,t_{0}),C_{2}(t,t_{0})]^{\mathcal{T}}$ is a two-component vector probability amplitude while $\mathbf{H}(t)$ is the total NH Hamiltonian of the system (with $\mathcal{T}$ denoting the transposed matrix). Irreversible decay of the system outside is analytically accounted for in $\mathbf{H}(t)$ by inserting negative imaginary terms $-i\Gamma_{1,2}$ in the diagonal part of the Hamiltonian $\mathbf{H}(t)$ which reads as follows:
\begin{eqnarray}\label{equ2}
\mathbf{H}(t)=\frac{1}{2}
\left[ {\begin{array}{*{20}c}
\Omega(t)-i\Gamma_{1} & \Delta(t)\\
 \Delta(t) & -\Omega(t)-i\Gamma_{2}
\end{array} } \right].
\end{eqnarray}
Amongst other things, the Hamiltonian $\mathbf{H}(t)$ is quite rich compared with the ones in Refs.~[\onlinecite{Vitanov1997, Avishai2014}] as not only one diabatic level decays, but all. We obtain exact analytical solutions to equation  (\ref{equ1}) for special selections of $\Omega(t)$ and $\Delta(t)$. We consider an experimentally useful choice, in which, the detuning is bounded at $t=\pm\infty$, smoothly depending on time and saturating at $T>0$ (pulse width). The detuning is made up of a time-dependent hyperbolic-tangent chirp and a static part (see also figure \ref{FIG1}),
\begin{eqnarray}\label{equ3}
\Omega(t)=
\Omega_{0}\tanh\Big(\frac{t}{T}\Big)+D,  
\end{eqnarray}
where $\Omega_{0}$ is the saturation energy at zero static detuning and $D$ the real part of the static detuning. This  selection gives us a wide range of possibilities to select the Rabi frequency $\Delta(t)$. We will consider two important cases: firstly, the case when $\Delta(t)$ assumes a time-dependent hyperbolic-secant shape $\Delta(t)=\Delta_{0}{\rm sech}(t/T)$ (decaying DK1 model) and secondly, the case when it is constant in time; $\Delta(t)=\Delta_{0}$ (decaying DK2 model).  Several special and interesting decaying models may be  analyzed from the decaying DK1 model. Namely, when $\Omega_{0}=0$, one deals with a decaying Rosen-Zener~\cite{Rosen} model while $D=0$ and $\Omega_{0}=D$ respectively correspond to decaying Bambini-Berman~\cite{Bambini} and Allen-Eberly~\cite{Allen} models.

In this paper, we are mostly interested in non-adiabatic evolutions realized when the system follows an eigenstate of the Hamiltonian in the absence of coupling. It is relevant to emphasize for further purposes that, eigen-energies associated with adiabatic states in the absence of decay, cross when $\Omega_{0}\ge D$ and do not in the opposite case $\Omega_{0}<D$ (see~\onlinecite{Nakamura} for ample discussions) i.e. globally,
\begin{eqnarray}\label{equa3a}
\left\{
{\begin{array}{*{20}c}
\hspace{-0.5cm}\Omega_{0}\ge D, \quad {\rm crossing\hspace{0.2cm} condition},\\\\
\Omega_{0}< D, \quad {\rm no-crossing\hspace{0.2cm} condition}.
\end{array} 
} \right.
\end{eqnarray}
In the absence of decay and $D\neq0$, the crossing condition $\Omega_{0}\ge D$ in equation  (\ref{equa3a}) is necessary and sufficient~\cite{Nakamura} (obviously when $D=0$, the  condition systematically holds).  We will demonstrate that in the presence of decay, this condition is just necessary but no longer sufficient. An additional condition, established in section \ref{Sec2.2} should be satisfied for occurrence of levels crossing. 

\subsection{Complex eigen-energies}\label{Sec2.2}

\begin{figure}[!h]
\vspace{-0.0cm}
 \begin{center}
          \includegraphics[width=4.2cm, height=40mm]{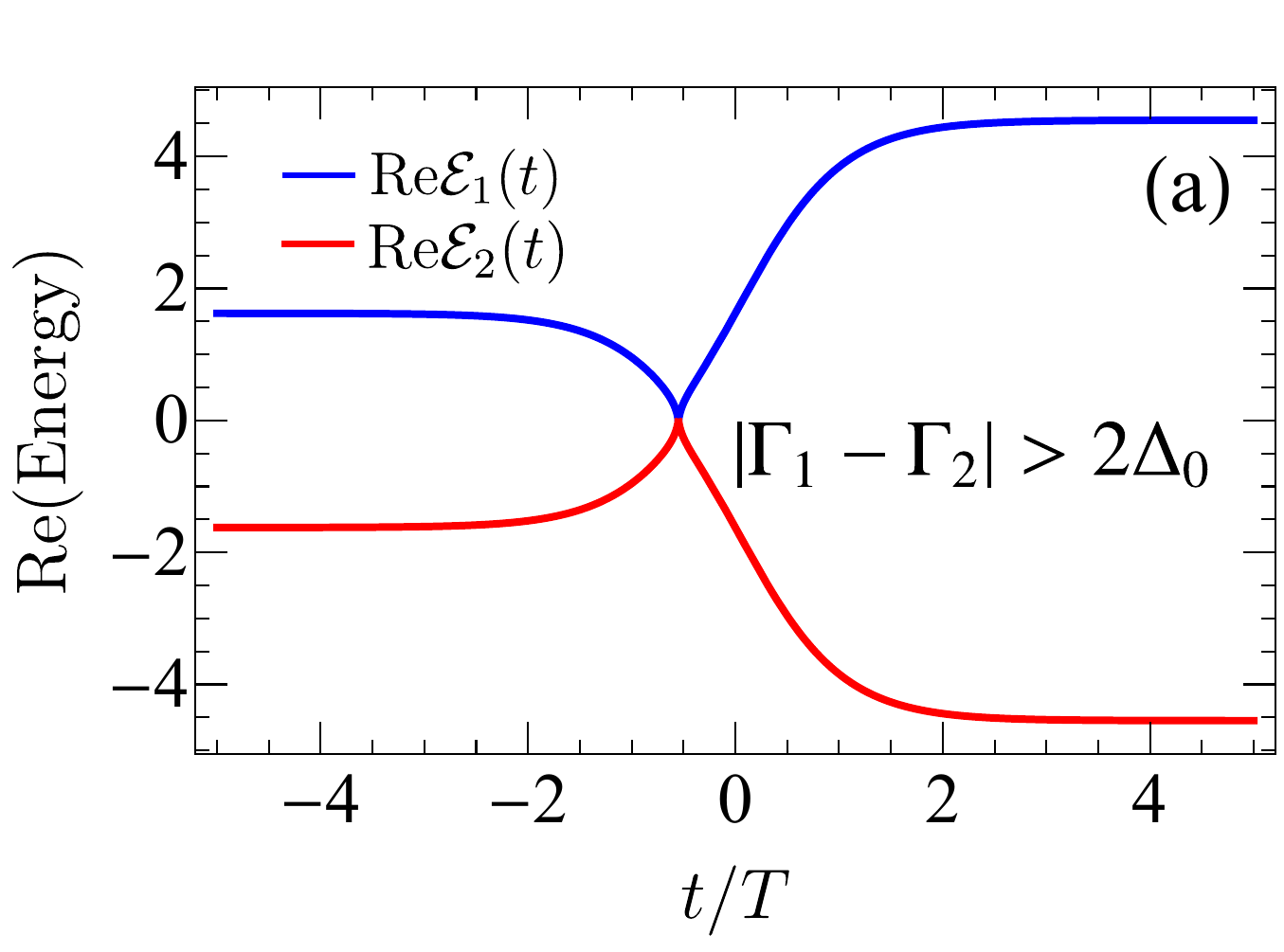}\hspace{-0.1cm}
					\includegraphics[width=4.2cm, height=40mm]{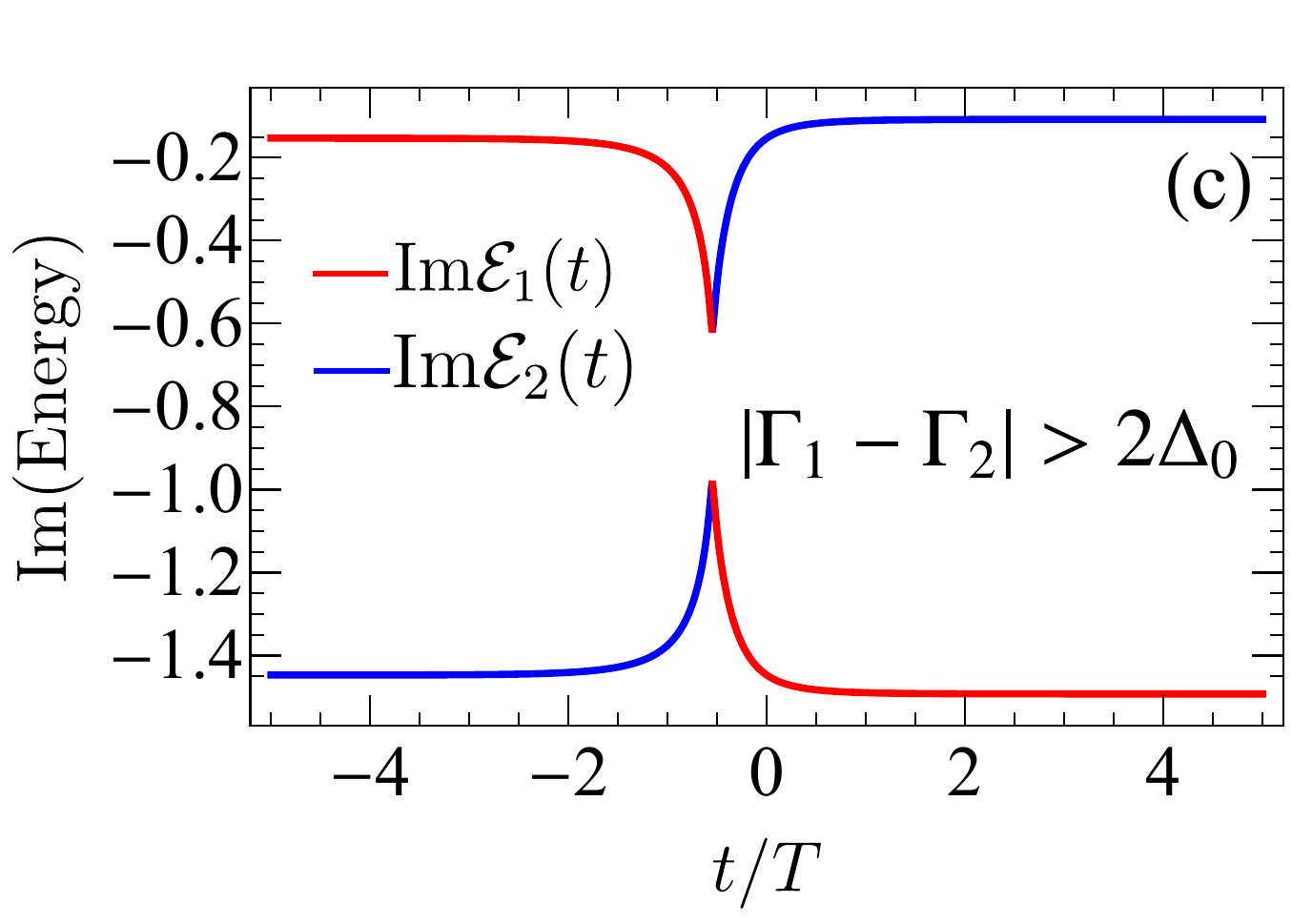}\\\vspace{-0.1cm}
					\includegraphics[width=4.2cm, height=40mm]{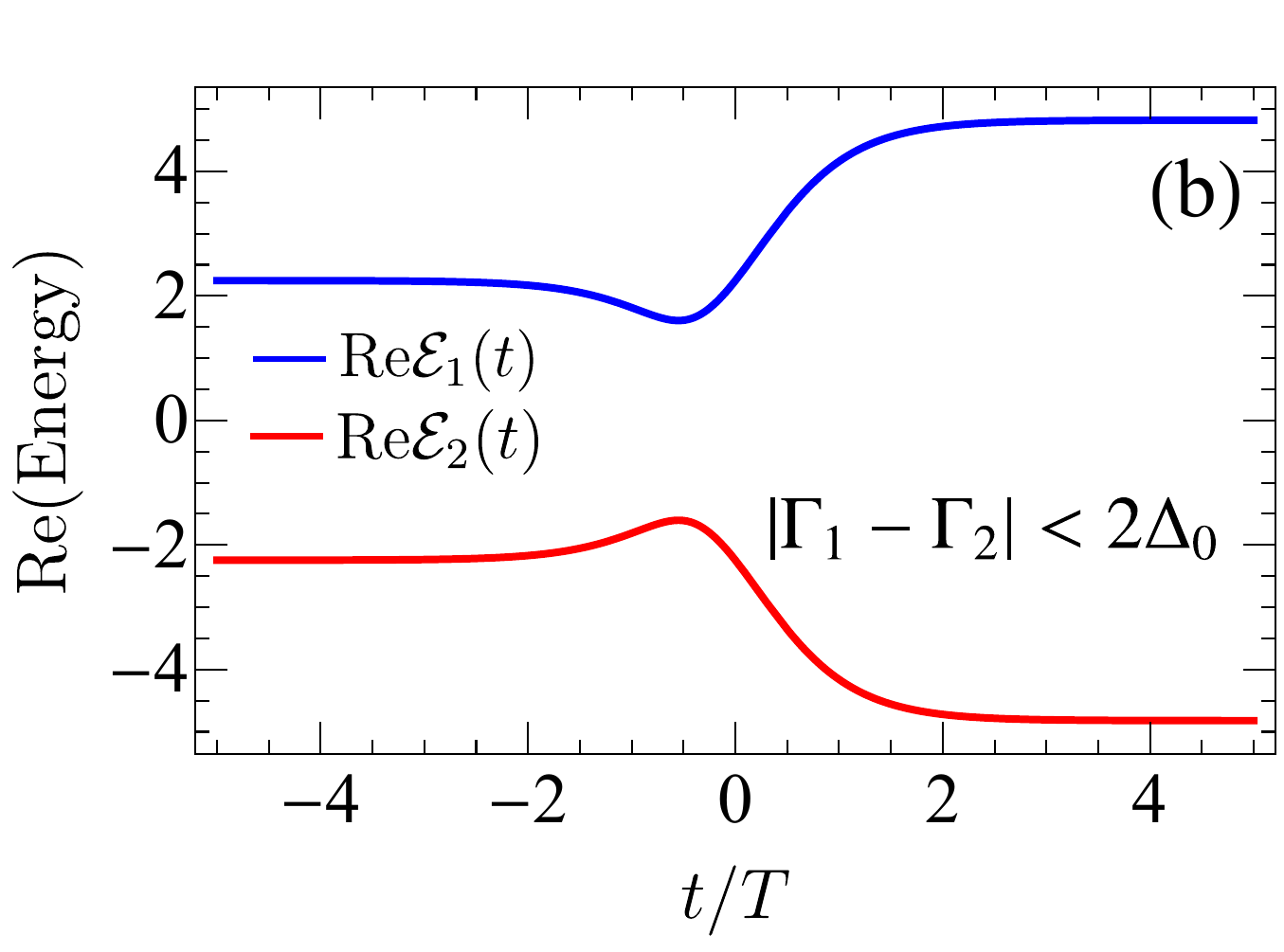}\hspace{-0.1cm} 
					\includegraphics[width=4.2cm, height=40mm]{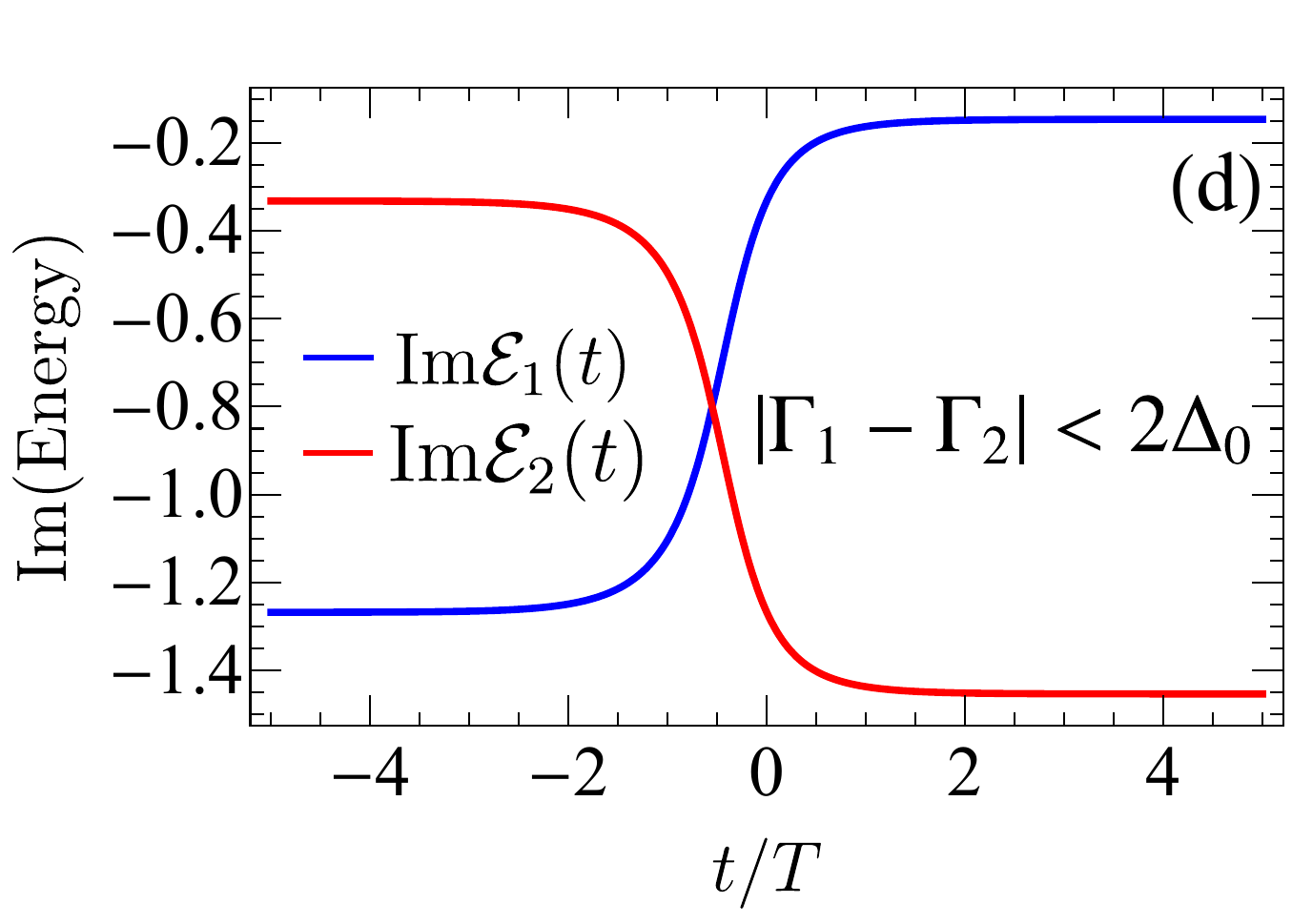}
\end{center}
\vspace{-0.7cm}
\caption{ Check of the crossing condition (\ref{equ3g}) for the DK2 model. On the panels (a) and (c), the condition (\ref{equ3g}) is respected ($\Omega_{0}=6/T$, $\Delta_{0}=1.35/T$, $D=3/T$, $\Gamma_{1}=0.2/T$ and $\Gamma_{2}=3/T$) and one observes a crossing of real parts of eigen-energies. On the panels (b) and (d), the condition is violated ($\Omega_{0}=6/T$, $\Delta_{0}=3.5/T$, $D=3/T$, $\Gamma_{1}=0.2/T$ and $\Gamma_{2}=3/T$) and the imaginary parts of eigen-energies cross.} \label{Comp}
\end{figure}

In order to put a borderline between both DK models and improve our theoretical investigations, we  analyze the eigen-energies
\begin{eqnarray}\label{equ3a}
\mathcal{E}_{1,2}(t)=-\frac{i}{2}\Big(\bar{\gamma}\pm i\Delta(t)\csc2\vartheta_{\bar{\Gamma}}(t)\Big),
\end{eqnarray}
of the Hamiltonian (\ref{equ2}). This allows us to capture many of its essential features. Here, we have denoted as $\bar{\Gamma}=(\Gamma_{1}-\Gamma_{2})/2$ and $\bar{\gamma}=(\Gamma_{1}+\Gamma_{2})/2$ and the time-dependent mixing angle $\vartheta_{\bar{\Gamma}}(t)$ obeys the relation
\begin{eqnarray}\label{equ3b}
\tan2\vartheta_{\bar{\Gamma}}(t)=-\frac{\Delta(t)}{\Omega(t)-i\bar{\Gamma}}.
\end{eqnarray}
$\bar{\Gamma}$ is introduced in equations  (\ref{equ3a}) and (\ref{equ3b}) to distinguish between the eigen-states of  $\mathbf{H}(t)$ in the absence and presence of decay. Due to decay, energies (\ref{equ3a}) can numerically be analyzed only in complex plane. Thus,
\begin{eqnarray}\label{equ3c}
\nonumber{\rm Re}\mathcal{E}_{1,2}(t)&=&\pm\frac{1}{2}\mathcal{W}(t)\cos\varphi(t)\\&=&
\pm\frac{1}{2}\sqrt{\bar{\Gamma}\Omega(t)\cot\varphi(t)},
\end{eqnarray}
and
\begin{eqnarray}\label{equ3d}
\nonumber {\rm Im}\mathcal{E}_{1,2}(t)&=&-\frac{1}{2}\Big[\bar{\gamma}\mp\mathcal{W}(t)\sin\varphi(t)\Big]\\&=&-\frac{1}{2}\Big[\bar{\gamma}\mp\sqrt{\bar{\Gamma}\Omega(t)\tan\varphi(t)}\Big],
\end{eqnarray}
are respectively the real and imaginary parts of $\mathcal{E}_{1,2}(t)$. Here, the function $\mathcal{W}(t)$ is defined  as 
\begin{eqnarray}\label{equ3f}
\nonumber\mathcal{W}(t)&=&\sqrt{\sqrt{\Big(\Omega^{2}(t)+\Delta^{2}(t)-\bar{\Gamma}^{2}\Big)^{2}+4\bar{\Gamma}^{2}\Omega^{2}(t)}}\\&=&\sqrt{2\bar{\Gamma}\Omega(t)\csc2\varphi(t)},
\end{eqnarray}
and the angle $\varphi(t)$ by
\begin{eqnarray}\label{equ3e}
\tan2\varphi(t)=\frac{2\Omega(t)\bar{\Gamma}}{\Omega^{2}(t)+\Delta^{2}(t)-\bar{\Gamma}^{2}}.
\end{eqnarray}
${\rm Re}\mathcal{E}_{1}(t)$ and ${\rm Re}\mathcal{E}_{2}(t)$ or ${\rm Im}\mathcal{E}_{1}(t)$ and ${\rm Im}\mathcal{E}_{2}(t)$ cross at a pseudo-crossing point $t_{cr}=T{\rm arctanh}(-D/\Omega_{0})$ defined such that $\Omega(t_{cr})=0$ and/or ${\rm Re}\mathcal{E}_{1}(t_{cr})-{\rm Re}\mathcal{E}_{2}(t_{cr})=0$. Thus, $\varphi(t_{cr})=0$ and $\mathcal{W}(t_{cr})=[\Delta^{2}(t_{cr})-\bar{\Gamma}^{2}]^{1/2}$. Interestingly, ${\rm Re}\mathcal{E}_{1,2}(t_{cr})=\pm\mathcal{W}(t_{cr})/2$ and ${\rm Im}\mathcal{E}_{1,2}(t_{cr})=-\bar{\gamma}/2$. It becomes evident from here that crossing of real/imaginary parts cannot only be attributed to the crossing condition in equation  (\ref{equa3a}) but also to decay of diabatic states. This fact is numerically confirmed (see fig.\ref{Comp}).  Thereof, in the presence of decay when $\Omega_{0}\ge D$,  the real parts of instantaneous complex energies cross at time $t_{cr}$ if the energy difference (gap) $\Delta(t_{cr})=(\mathcal{E}_{1}(t_{cr})-\mathcal{E}_{2}(t_{cr}))|_{\Gamma_{1}=\Gamma_{2}=0}$ between adiabatic states in the  absence of decay is smaller than (or equal to) the half of decay rates difference (namely $\bar{\Gamma}$) in absolute value. Thence, the condition for occurrence of level crossing in addition to $\Omega_{0}\ge D$  reads
\begin{eqnarray}\label{equ3g}
|\Gamma_{1}-\Gamma_{2}|\ge2\Delta(t_{cr}).  
\end{eqnarray}
When the condition (\ref{equ3g}) is satisfied, the real parts of eigenvalues cross while imaginary parts do not [see figures \ref{Comp}(a) and \ref{Comp}(c)]. When this condition is violated, imaginary parts now cross and real parts do not [see figures \ref{Comp}(b) and \ref{Comp}(d)]. The condition $\Omega_{0}\ge D$ is thus necessary but not sufficient.

For obvious reasons and for DK1, $\Delta(t_{cr})=\Delta_{0}\sqrt{1-D^{2}/\Omega_{0}^{2}}$ and for the DK2 model, $\Delta(t_{cr})=\Delta_{0}$. Importantly, the point $t_{cr}$ can be located either at the left or right of the point $t=0$ (crossing point in the absence of static detuning) or exactly at that point depending on how strong $D$ is. When $D>0$, the crossing time $t_{cr}<0$ and is located at the left of $t=0$ when in contrary $D<0$, the point $t_{cr}>0$ is located at the right. For the generic case $D=0$, then $t_{cr}=0$. The real part of the static detuning shifts the pseudo-crossing point $t_{cr}=0$ from a value $T{\rm arctanh}(-D/\Omega_{0})$ and may be considered as a control parameter~\cite{Vitanov2007}. 

\section{Theory}\label{Sec3}
For both cases discussed in this paper (decaying DK1 and DK2 models), the dynamics of the system is encoded into the time-dependent Schr\"odinger equation (\ref{equ1}). The central goal is to determine the probability amplitudes $C_{1,2}(t,t_{0})$. They obey the same family of second-order differential equations (as shown below) and their resolutions follow the same mathematical procedure. We have thus found opportune to construct a general theory which applies to both models.  

\subsection{Dynamical phases extraction}\label{Sec3.1}

As there is a loss of probability, the total wave function is disentangled and cannot be expressed as a linear superposition of the subsystems' states. Therefore, it is instructive to modulate the probability amplitudes through the gauge transformations
\begin{eqnarray}\label{equ4}
C_{1,2}(t,t_{0})=\Psi_{1,2}(t)\exp\Big(-i\Phi_{1,2}(t,t_{0})\Big).
\end{eqnarray}
This extracts contributions of dynamical phases picked-up by the system during the rising phase of the pulse. This also leads to a simpler problem for $\Psi_{1,2}(t)$ and do not affect the total population. Here, the characteristic oscillatory phases 
\begin{eqnarray}\label{equ5}
\Phi_{1,2}(t,t_{0})=\int_{t_{0}}^{t}dt'\delta_{1,2}(t'),
\end{eqnarray}
are dynamical phases acquired by each of the two components of the total wave function  during adiabatic stages and $\delta_{1,2}(t)=(\pm\Omega(t)-i\Gamma_{1,2})/2$, the complex diabatic energies of the total Hamiltonian associated with the bare states $|1\rangle$  and $|2\rangle$. For completeness, we introduce the energy difference $\delta_{12}(t)=\delta_{1}(t)-\delta_{2}(t)$ where non-adiabatic transitions occur in the region $\delta_{12}(t)=0$ for hermitian Hamiltonians. This in general cannot be realized with NH Hamiltonians because of decay~\cite{Avishai2014}. Then, the phases (\ref{equ5}) are separated in two relevant contributions and rewritten as
\begin{eqnarray}\label{equ6}
\Phi_{1,2}(t,t_{0})=\Phi_{1,2}^{\rm (pulse)}(t,t_{0})+\Phi_{1,2}^{\rm (decay)}(t,t_{0}).
\end{eqnarray}
The seek for approximated solutions (semi-classical approach for instance) leads in general to third terms $\Phi_{1,2}^{\rm (geom)}(t,t_{0})$ (geometric phase) entering  the description of the system's dynamics.  The contribution of these phases to non-adiabatic transfer processes with decay was revealed in ~ Ref.\onlinecite{Garanin}. Here, they do not play any role as we are searching for exact analytical solutions. The first terms in equation  (\ref{equ6}) are phases accumulated during the sweep of the external pulse. They usually converge and create oscillations in the  population.  On the other hand, the phases $\Phi_{1,2}^{\rm (decay)}(t,t_{0})$ are imaginary and create exponential decrease/decay of the total population such that when $\Delta(t)=0$, populations on diabatic states vanish  when $t=+\infty$. 

\subsection{Reduction to transcendental equations}\label{Sec3.2}

The functions $\Psi_{1,2}(t)$ in equation  (\ref{equ4}) do not explicitly depend on $t_{0}$ and satisfy the linear second-order differential equations
\begin{eqnarray}\label{equ7}
\frac{d^{2}\Psi_{1,2}}{dt^{2}}-\Big(\frac{\dot{\Delta}(t)}{\Delta(t)}\pm i\delta_{12}(t)\Big)\frac{d\Psi_{1,2}}{dt}+\frac{\Delta^{2}(t)}{4}\Psi_{1,2}=0.\quad
\end{eqnarray}
The dots on functions denote time derivatives. Under the present form, the equation (\ref{equ7}) is not easily handled. For each of the two cases faced in this paper (hyperbolic-secant and constant Rabi frequencies), a unique and single change of variable is used~\cite{Garraway}
\begin{eqnarray}\label{equ8}
z(t)=\frac{1}{2}\Big(1+\tanh\Big(\frac{t}{T}\Big)\Big).
\end{eqnarray}
Our approach consists of first determining the function $\Psi_{1}(z)$. Through the change of variable (\ref{equ8}), the equation (\ref{equ7}) for $\Psi_{1}(z)$ acquires the form
\begin{eqnarray}\label{equ9}
 z(1-z)\frac{\partial^{2}\Psi_{1}}{\partial z^{2}}+(a-bz)\frac{\partial\Psi_{1}}{\partial z}+c^{2}\mathcal{R}(z)\Psi_{1}=0,
\end{eqnarray}
where the parameters $a$, $b$ and $c$ as well as the function $\mathcal{R}(z)$ are problem-dependent. They are presented in section \ref{Sec4a} for DK1 and in section \ref{Sec4b} for DK2. Thus, the equation (\ref{equ7}) for $\Psi_{1}(z)$ is transformed to a higher transcendental equation for special functions  with the aid of the anzath
\begin{eqnarray}\label{equ10}
\Psi_{1}(z)=z^{\mu}(1-z)^{\nu}\mathcal{Y}(z),
\end{eqnarray}
where $\mu$ and $\nu$ are presented in sections \ref{Sec4a} and \ref{Sec4b} for DK1  and DK2 respectively. As it will be seen further in this work, they strongly depend on the shape of $\mathcal{R}(z)$. The function $\mathcal{Y}(z)$ in equation  (\ref{equ10}) satisfies the Gauss hypergeometric equation~\cite{MathBook, Wong}
\begin{eqnarray}\label{equ11}
z(1-z)\frac{\partial^{2}\mathcal{Y}}{\partial z^{2}}+[\gamma-(\alpha+\beta+1)z]\frac{\partial\mathcal{Y}}{\partial z}-\alpha\beta\mathcal{Y}=0.\quad
\end{eqnarray}
Here, $\alpha$, $\beta$ and $\gamma$ are functions of $\mu$ and $\nu$ (see sections \ref{Sec4a} and \ref{Sec4b}). Equation (\ref{equ11}) possesses two linearly independent solutions $F(\alpha, \beta, \gamma; z)$ and $z^{1-\gamma}F(\alpha-\gamma+1, \beta-\gamma+1, 2-\gamma; z)$ (where $F(...)$ is the Gauss hypergeometric function)~\cite{MathBook, Wong}. Thus, the function $\mathcal{Y}(z)$ is constructed as a linear superposition of these two solutions providing two constants that are functions of initial preparation of the system.
 
 To return to equation  (\ref{equ4}) and construct $C_{1}(t,t_{0})$ as well as $C_{2}(t,t_{0})$, one needs the exponential phase 
\begin{eqnarray}\label{equ11a}
\exp\Big[-i\Phi_{1}(z,z_{0})\Big]=z^{\lambda}(1-z)^{\eta}z^{-\lambda}_{0}(1-z_{0})^{-\eta},
\end{eqnarray}
where
\begin{eqnarray}\label{equ11b}
\lambda=\frac{i\Omega_{0}T}{4}-\frac{\Gamma_{1}T}{4}-\frac{iDT}{4}, 
\end{eqnarray}
\begin{eqnarray}\label{equ11bb}
\eta=\frac{i\Omega_{0}T}{4}+\frac{\Gamma_{1}T}{4}+\frac{iDT}{4}.
\end{eqnarray}
Thus, one sees that the expected exponential decay observed in $t$-space is renormalized by polynomials raised to the powers $\eta$ and $\lambda$ in the $z$-space. The fact that only $\Gamma_{1}$ enters  $\lambda$ and $\eta$ is explained by the choice we have made to first evaluate $\Psi_{1}(z)$. The situation would be different (i.e only $\Gamma_{2}$ in  $\lambda$ and $\eta$) if one rather chooses to first evaluate $\Psi_{2}(z)$. Finally, one obtains
\begin{eqnarray}\label{equ155}
C_{1}(z,z_{0})=z^{\mu+\lambda}(1-z)^{\nu+\eta}\mathcal{Y}(z)z^{-\lambda}_{0}(1-z_{0})^{-\eta}.
\end{eqnarray}
The function $\Psi_{2}(z)$ is deduced from the Schr\"odinger equation (\ref{equ1}) using derivative properties (\ref{02}) and (\ref{03}) of Gauss hypergeometric functions. At the end, 
\begin{eqnarray}\label{equ15}
\nonumber C_{1,2}(z,z_{0})=z^{\lambda}(1-z)^{\eta}\Big(a_{+}(z_{0})\mathsf{U}_{1,2}(z)+\mathsf{V}_{1,2}(z)a_{-}(z_{0})\Big),\\
\end{eqnarray}
where
\begin{eqnarray}\label{equ17}
\nonumber a_{+}(z_{0})=\frac{C_{1}(z_{0},z_{0})\mathsf{V}_{2}(z_{0})-\mathsf{V}_{1}(z_{0})C_{2}(z_{0},z_{0})}{\mathsf{U}_{1}(z_{0})\mathsf{V}_{2}(z_{0})-\mathsf{V}_{1}(z_{0})\mathsf{U}_{2}(z_{0})}z^{-\lambda}_{0}(1-z_{0})^{-\eta},\\
\end{eqnarray}
\begin{eqnarray}\label{equ18}
\nonumber a_{-}(z_{0})=\frac{C_{2}(z_{0},z_{0})\mathsf{U}_{1}(z_{0})-\mathsf{U}_{2}(z_{0})C_{1}(z_{0},z_{0})}{\mathsf{U}_{1}(z_{0})\mathsf{V}_{2}(z_{0})-\mathsf{V}_{1}(z_{0})\mathsf{U}_{2}(z_{0})}z^{-\lambda}_{0}(1-z_{0})^{-\eta}.\\
\end{eqnarray}
Here, $C_{1}(z_{0},z_{0})$ and $C_{2}(z_{0},z_{0})$ are probability amplitudes at initial time $t_{0}$. The functions $\mathsf{U}_{1,2}(z)$ and $\mathsf{V}_{1,2}(z)$ are expressed in terms of hypergeometric functions and presented in appendix \ref{App2}.  Equations  (\ref{equ15})-(\ref{equ18}) are general solutions to our problems and are used to compute transition probabilities and the full propagator. If the system starts at time $t_{0}$ in the diabatic state $|\kappa\rangle$, the initial conditions read $C_{j}(z_{0},z_{0})=\delta_{j\kappa}$ with $(j,\kappa$)=1,2. When in addition $j=\kappa$, the function $P_{j}(z,z_{0})=|C_{j}(z,z_{0})|^{2}$, is the population which returns to the diabatic state $|\kappa\rangle$ after interactions while for $j\neq\kappa$, the function $P_{j}(z,z_{0})$ describes the population transferred to the excited-state. In order to describe the complete evolution of the system, it will be relevant to evaluate its propagator. 

\subsection{Propagator}\label{Sec3.3}

We construct the full propagator $\mathbf{U}(z,z_{0})$ describing the system evolution from the initial time $t_{0}$ to an arbitrary time $t$ and connecting the vectors probability amplitudes as $\mathbf{C}(z)=\mathbf{U}(z,z_{0})\mathbf{C}(z_{0})$. The elements of $\mathbf{U}(z,z_{0})$ are calculated from equations  (\ref{A5})-(\ref{A6}) using (\ref{04}). Letting $U_{\kappa\kappa'}(z,z_{0})\equiv U_{\kappa\kappa'}^{(\Gamma_{1}, \Gamma_{2})}(z,z_{0})$, as results, one obtains,
\begin{widetext}
\begin{eqnarray}\label{equ19}
\Big[U_{11}^{(\Gamma_{2}, \Gamma_{1})}(z,z_{0})\Big]^{*}=U_{22}^{(\Gamma_{1}, \Gamma_{2})}(z,z_{0})=\frac{\Gamma(\alpha)\Gamma(\beta)e^{\mathcal{V}(z,z_{0})}}{\Gamma(\alpha+\beta+1-\gamma)\Gamma(\gamma)}\frac{d}{dz}\Big[z^{\mu}(1-z)^{\nu}\Big(G_{\alpha\beta}^{\gamma}(z,z_{0})-G_{\alpha\beta}^{\gamma}(z_{0},z)\Big)\Big],
\end{eqnarray}

and
\begin{eqnarray}\label{equ20}
U_{12}^{(\Gamma_{1}, \Gamma_{2})}(z,z_{0})=-\Big[U_{21}^{(\Gamma_{2}, \Gamma_{1})}(z,z_{0})\Big]^{*}=\frac{ic_{0}\Gamma(\alpha)\Gamma(\beta)}{\Gamma(\alpha+\beta+1-\gamma)\Gamma(\gamma)}\Big[G_{\alpha\beta}^{\gamma}(z,z_{0})-G_{\alpha\beta}^{\gamma}(z_{0},z)\Big]e^{\vartheta(z,z_{0})},
\end{eqnarray}
\end{widetext}
where $\Gamma(...)$ denotes the Euler's Gamma function~\cite{MathBook, Wong}. The phases $\mathcal{V}(z,z_{0})$ and $\vartheta(z,z_{0})$ are given by the relations
\begin{eqnarray}\label{equ2.12ee}
\mathcal{V}(z,z_{0})=\vartheta(z,z_{0})+(\theta-\mu)\ln z+(\theta-\nu)\ln(1-z),\quad
\end{eqnarray}
and
\begin{eqnarray}\label{equ2.12d}
\nonumber\vartheta(z,z_{0})=(\lambda+\mu)\ln z\\\nonumber+(\eta+\nu)\ln(1-z)-(\lambda+\mu+\theta-\gamma)\ln z_{0}\\-(\eta+\nu+\gamma-\alpha-\beta-1+
\theta)\ln(1-z_{0}).
\end{eqnarray}
Similarly,
\begin{eqnarray}\label{equ2.12c}
G_{\alpha\beta}^{\gamma}(z,z_{0})=F(\alpha,\beta,\gamma; z)F(\alpha,\beta,\alpha+\beta-\gamma+1; 1-z_{0}).\qquad
\end{eqnarray}
The stars on functions in equations  (\ref{equ19}) and (\ref{equ20}) indicate complex conjugate. Note that for the DK1 model, $c_{0}=c$ and $\theta=1$ while for DK2, $c_{0}=2c$ and $\theta=1/2$ ($c$ is given below for each model). For further purposes and for the sake of completeness, using the property (\ref{09}), it  can be shown that
\begin{eqnarray}\label{equ2.12e}
G_{\alpha\beta}^{\gamma}(1,0)=\frac{\sin[\pi(\gamma-\alpha)]\sin[\pi(\gamma-\beta)]}{\sin[\pi(\gamma)]\sin[\pi(\gamma-\alpha-\beta)]},
\end{eqnarray}
and
\begin{eqnarray}\label{equ2.12f}
G_{\alpha\beta}^{\gamma}(1,0)-G_{\alpha\beta}^{\gamma}(0,1)=\frac{\sin[\pi\alpha]\sin[\pi\beta]}{\sin[\pi\gamma]\sin[\pi(\gamma-\alpha-\beta)]}.\quad
\end{eqnarray}
Some derivative properties of the function $G_{\alpha\beta}^{\gamma}(z,z_{0})$ are given below. Let us define
\begin{eqnarray}\label{equ2.12h}
Q_{\alpha\beta}^{\gamma}(z,z_{0})=\frac{d}{dz}\Big[z^{\gamma-1}(1-z)^{\alpha+\beta-\gamma}G_{\alpha\beta}^{\gamma}(z,z_{0})
\Big].
\end{eqnarray} 
Thus,
\begin{eqnarray}\label{equ2.12g}
\nonumber Q_{\alpha\beta}^{\gamma}(z,z_{0})=(\gamma-1)z^{\gamma-2}F(\gamma-\alpha, \gamma-\beta, \gamma-1; z)\\\times F(\alpha, \beta, \alpha+\beta-\gamma+1; 1-z_{0}).
\end{eqnarray} 
Similarly, 
\begin{eqnarray}\label{equ2.12h}
\bar{Q}_{\alpha\beta}^{\gamma}(z_{0},z)=\frac{d}{dz}\Big[z^{\gamma-1}(1-z)^{\alpha+\beta-\gamma}G_{\alpha\beta}^{\gamma}(z_{0},z)
\Big],
\end{eqnarray} 
and we find that
\begin{eqnarray}\label{equ2.12i}
\nonumber\bar{Q}_{\alpha\beta}^{\gamma}(z_{0},z)=(\gamma-\alpha-\beta)(1-z)^{\alpha+\beta-\gamma-1}F(\alpha, \beta,\gamma; z_{0})\\\times F(1+\alpha-\gamma, 1+\beta-\gamma, \alpha+\beta-\gamma; 1-z).\qquad
\end{eqnarray}
These relations are helpful to evaluate the propagator component (\ref{equ19}) and consequently occupation and transition probabilities.

For arbitrary decay rates including $\Gamma_{1,2}=0$, equations  (\ref{equ19}) and (\ref{equ20}) assert that $U_{11}^{(\Gamma_{1}, \Gamma_{2})}(z,z_{0})=U_{22}^{(\Gamma_{1}, \Gamma_{2})}(z_{0},z)$. The evolution matrix is not unitary at a given time 
$t>t_{0}$. This is due to decay of diabatic states i.e. $|U_{11}^{(\Gamma_{1}, \Gamma_{2})}(z\ge z_{0},z_{0})|^{2}+|U_{21}^{(\Gamma_{1}, \Gamma_{2})}(z\ge z_{0},z_{0})|^{2}\le1$ and there is a loss of probability. On the other hand, the same equations suggest that $U_{11}^{(\Gamma_{1}, \Gamma_{2})}(z_{0},z_{0})=U_{22}^{(\Gamma_{1}, \Gamma_{2})}(z_{0},z_{0})=1$ and $U_{21}^{(\Gamma_{1}, \Gamma_{2})}(z_{0},z_{0})=U_{12}^{(\Gamma_{1}, \Gamma_{2})}(z_{0},z_{0})=0$ i.e. the propagator is unitary at least  at initial time $t_{0}$ when the system is switched on and this unitary character is lost when the evolution becomes effective in time. Our representations however indicate that the remaining parts of the propagator are deduced by swapping $\Gamma_{1}\rightleftharpoons\Gamma_{2}$ everywhere they appear and performing further operations. The full propagator obtained is useful to describe the system during cyclic evolutions (periodic drive for instance). 

\subsection{Transition probability}\label{sec3.4}

The probability of transition between the diabatic states $|1\rangle$ and $|\kappa\rangle$ is calculated from $P_{1\to\kappa}(z,z_{0})=|U_{\kappa1}^{(\Gamma_{1}, \Gamma_{2})}(z,z_{0})|^{2}$ and $P_{\kappa\to1}(z,z_{0})=|U_{1\kappa}^{(\Gamma_{1}, \Gamma_{2})}(z,z_{0})|^{2}$. Let us denote by $P_{\kappa}(z,z_{0})\equiv P_{1\to\kappa}(z,z_{0})$. Therefore, considering the relations  (\ref{equ19}) and (\ref{equ20}), $P_{1}(z,z_{0})\equiv|U_{22}^{(\Gamma_{2}, \Gamma_{1})}(z,z_{0})|^{2}$ and $P_{2}(z,z_{0})\equiv|U_{12}^{(\Gamma_{2}, \Gamma_{1})}(z,z_{0})|^{2}$ are respectively the survival probability of the state $\psi_{1}(z,z_{0})$ and the probability of transition to the state $\psi_{2}(z,z_{0})$ at a given instant $t$ if the system starts at time $t_{0}$ in the state $\psi_{1}(z,z_{0})$. In what follows, we present the excited-state probability 
\begin{eqnarray}\label{equ28}
\nonumber P_{2}(z,z_{0})\hspace{7cm}\\=\Big|\frac{c_{0}\Gamma(\alpha)\Gamma(\beta)e^{\vartheta(z,z_{0})}}{\Gamma(1-\gamma+\alpha+\beta)\Gamma(\gamma)}\Big[G_{\alpha\beta}^{\gamma}(z,z_{0})-G_{\alpha\beta}^{\gamma}(z_{0},z)\Big]\Big|^{2}.\quad
\end{eqnarray}
This solution is identical for DK1 and DK2. According to previous discussions, the swap $\Gamma_{1}\leftrightarrow\Gamma_{2}$ should be operated in equation  (\ref{equ28}) to ensure that $P_{2}(z,z_{0})$ describes $P_{1\to2}(z,z_{0})$. This operation is equivalent to the change $\Gamma_{1}\to\Gamma_{2}$ in $\lambda$ and $\eta$. With our results, the effects of real and imaginary parts of the static detuning on transition probabilities can be investigated. The turn-on and turn-off times can be manipulated as well. Our solutions  are consequently generalizations and extensions (all levels decay) of results in Refs.~[\onlinecite{Vitanov1997, Avishai2014}]. Note that, because of decay  inducing non-unitary evolutions, the survival probability $P_{1}(z,z_{0})$ cannot be deduced from $P_{2}(z,z_{0})$ in equation (\ref{equ28}) but from (\ref{equ19}).

Consider the extremal limit $t_{0}=-\infty$ as the turn-on time and $t=+\infty$ as turn-off time. Thus, $z_{0}=0$ and $z=1$ respectively. Taking into account the properties of $G_{\alpha\beta}^{\gamma}(z,z_{0})$ in equations  (\ref{equ2.12e}) and (\ref{equ2.12f}), one obtains the large positive time solution
\begin{eqnarray}\label{equ29}
\nonumber P_{2}(1,0)\hspace{7cm}\\= 
\Big|\frac{c_{0}\Gamma(\alpha)\sin[\pi\alpha]\Gamma(\beta)\sin[\pi\beta]e^{\vartheta(1,0)}}{\Gamma(\alpha+\beta+1-\gamma)\sin[\pi(\gamma-\alpha-\beta)]\Gamma(\gamma)\sin[\pi\gamma]}\Big|^{2}.\quad
\end{eqnarray}
This expression is valid when $\Gamma_{1}$ and $\Gamma_{2}$ are selected such that ${\rm Re}(\gamma)>{\rm Re}(\alpha+\beta)$ [limitation inherent to the property (\ref{07})]. When $\Gamma_{1,2}=0$, equation  (\ref{equ29}) yields the exact results in~Refs.[\onlinecite{ Garraway, Nakamura}].

\section{First and second DK models with decay}\label{Sec4.0}
\subsection{First DK model with decay}\label{Sec4a}

Here, we illustrate the general theory presented above by the most famous version of DK models: the first DK model (DK1). In the prototype Hamiltonian,  interactions are regulated by a time-dependent hyperbolic-secant function
\begin{eqnarray}\label{equ4.1}
\Delta(t)=\Delta_{0}{\rm sech}\Big(\frac{t}{T}\Big).
\end{eqnarray}
If only the excited-state is allowed to decay, the problem reduces to the one studied in Ref.~\onlinecite{Vitanov1997}. As already pointed out, our analytical solution  
(\ref{equ28}) accounts not only for the initial time $t_{0}=-\infty$ as in Ref.~\onlinecite{Vitanov1997} but also for arbitrary initial time including $t_{0}=0$ and is more advantageous and useful for experiments.

For the model of interest in this section (decaying DK1 model), the function $\mathcal{R}(z)$ in equation  (\ref{equ9}) takes a relatively simple form 
\begin{eqnarray}\label{equ4.1a}
\mathcal{R}(z)=1,
\end{eqnarray}
and the parameters $a$, $b$ and $c$ which compose equation  (\ref{equ9}) are given by
\begin{eqnarray}\label{equ4.2}
a=\frac{1}{2}-\frac{i\Omega_{0}T}{2}+\frac{iDT}{2}+\frac{\bar{\Gamma} T}{2},
\end{eqnarray}
\begin{eqnarray}\label{equ4.2a}
b=1-i\Omega_{0}T,\quad {\rm and}\quad
c=\frac{\Delta_{0} T}{2}.
\end{eqnarray}
In the anzath equation  (\ref{equ10}), $\mu$ and $\nu$ are found by substituting equation  (\ref{equ10}) into equation  (\ref{equ9}). In the resulting equation, they are selected such that the remaining equation is of Gauss hypergeometric form~\cite{MathBook, Wong}. This leads to a pair of second order algebraic equations and yield two pairs of solutions, the first pair of which is trivial ($\mu=\nu=0$) and the second one considered here (non-trivial) is 
\begin{eqnarray}\label{equ4.3}
\mu=1-a, 
\end{eqnarray}
\begin{eqnarray}\label{equ4.3a}
\nu=1+a-b.
\end{eqnarray}
In the resulting higher transcendental equation (\ref{equ11}), the parameters $\alpha$, $\beta$ and $\gamma$ are given by 
\begin{eqnarray}\label{equ4.4}
\alpha=\frac{1}{2}\Big(3-b+\sqrt{(1-b)^{2}+4c^{2}}\Big),
\end{eqnarray}
\begin{eqnarray}\label{equ4.5}
\beta=\frac{1}{2}\Big(3-b-\sqrt{(1-b)^{2}+4c^{2}}\Big),
\end{eqnarray}
and
\begin{eqnarray}\label{equ4.6}
\gamma=2-a.
\end{eqnarray}
Having all the parameters (\ref{equ4.2})-(\ref{equ4.6}), the functions $\mathsf{U}_{1,2}(z)$ and $\mathsf{V}_{1,2}(z)$ deferred in Appendix \ref{App2} fully determine the total propagator and this achieves our goal. The latter is defined by equations  (\ref{equ19}) and (\ref{equ20}). It should be noted that the relations $\gamma-\alpha-\beta=-\nu$ and $1-\gamma=-\mu$ are useful to simplify the functions $\mathsf{U}_{1,2}(z)$ and $\mathsf{V}_{1,2}(z)$ using the property  (\ref{06}). 

\subsection{Second DK model with decay}\label{Sec4b}

The DK2 model is ruled by the Hamiltonian (\ref{equ2}) in which the detuning is given by equation  (\ref{equ3}) and the Rabi frequency (interaction) is constant:
\begin{eqnarray}\label{equ2.1}
\Delta(t)=\Delta_{0}.
\end{eqnarray}
When $D=0$ and only the excited-state is allowed to decay outside,  the relevant problem corresponds to the one discussed in Ref.~\onlinecite{Avishai2014}. In the slow rise regime $1/T\to0$, DK2 reduces to a decaying LZ model~\cite{Vitanov1997, Avishai2014, See, AkulinBook, Moyer, Garanin, Kenmoe2015}. In the complementary limit, $1/T\to \infty$ (fast rise) it yields a decaying Rabi model (see section \ref{Sec6}). 

Considering as suggested the change of variable (\ref{equ8}), we achieve an equation similar to equation  (\ref{equ9}) where the function $\mathcal{R}(z)$ explicitly reads
\begin{eqnarray}\label{equ2.2}
\mathcal{R}(z)=z^{-1}(1-z)^{-1},
\end{eqnarray}
and the parameters $a$, $b$ and $c$ are given by the expressions
\begin{eqnarray}\label{equ2.3}
a=1+\frac{i\Omega_{0}T}{2}-\frac{iDT}{2}-\frac{\bar{\Gamma} T}{2},
\end{eqnarray}
\begin{eqnarray}\label{equ2.3a}
b=2+i\Omega_{0}T, \quad {\rm and}\quad c=\frac{\Delta_{0} T}{4}.
\end{eqnarray}
Furthermore,  
\begin{eqnarray} 
\mu=\frac{1}{2}\Big(1-a-\sqrt{(1-a)^{2}-4c^{2}}\Big),\quad 
\end{eqnarray}
and
\begin{eqnarray}\label{equ2.4}
\nu=\frac{1}{2}\Big(1+a-b+\sqrt{(1+a-b)^{2}-4c^{2}}\Big).
\end{eqnarray}
The resulting hypergeometric equation is that of Gauss equation  (\ref{equ11}) where
\begin{eqnarray}\label{equ2.7a}
\nonumber\alpha=\frac{1}{2}\Big(2-b-\sqrt{(1-a)^{2}-4c^{2}}+\sqrt{(1+a-b)^{2}-4c^{2}}\Big),\\
\end{eqnarray}
\begin{eqnarray}\label{equ2.7b}
\nonumber\beta=\frac{1}{2}\Big(b-\sqrt{(1-a)^{2}-4c^{2}}+\sqrt{(1+a-b)^{2}-4c^{2}}\Big),\\
\end{eqnarray}
and
\begin{eqnarray}\label{equ2.7}
\gamma=1-\sqrt{(1-a)^{2}-4c^{2}}. 
\end{eqnarray}
 We have thus achieved the exact probability amplitudes (\ref{equ15})-(\ref{equ18}) where the functions $\mathsf{U}_{1,2}(z)$ and $\mathsf{V}_{1,2}(z)$ are presented in Appendix \ref{App2}. 
It should be noted that the following relations: $\alpha=\mu+\nu$, $\beta=\mu+\nu+b-1$ and $\gamma=2\mu+a$ may be used to rewrite our solutions. Another possible choice of $\mu$ and $\nu$ is given by $\mu=\frac{1}{2}[1-a+\sqrt{(1-a)^{2}-4c^{2}}]$ and $\nu=\frac{1}{2}[1+a-b-\sqrt{(1+a-b)^{2}-4c^{2}}]$. This yields a different set of parameters $\alpha$, $\beta$ and $\gamma$ (not shown here).

\section{Limiting cases}\label{Sec6}
\subsection{Fast rise, Rabi model}\label{Sec6.1}
In the fast rise limit, the frequency $1/T\to\infty$ as $T\to0$, the time-dependent part of the detuning turns to a step-like function. During the rising phase of the pulse in the time interval $]-\infty,0]$, it  saturates around $-1$, abruptly jumps at $t=0$ and saturates again at $+1$ in the time interval $[0,\infty[$,
\begin{eqnarray}\label{l1}
\tanh\Big(\frac{t}{T}\Big)\approx
\left\{
{\begin{array}{*{20}c}
1-\epsilon, \quad t\ge0,\\\\
\epsilon-1, \quad t<0
\end{array} 
} \right.
\end{eqnarray}
where $\epsilon=2e^{-2t/T}$ is a small parameter $0\le\epsilon\ll1$. The coupling in the DK1 model vanishes as ${\rm sech}(t/T)\to0$. Relevantly, the  DK1 model with time-dependent coupling cannot be reduced to a Rabi model. The  DK2 model reduces to the Rabi model and deserves attention due to multiple applications in quantum physics~\cite{Rabi}.  

\begin{figure}[!h]
 \begin{center}
          \includegraphics[width=8.5cm, height=65mm]{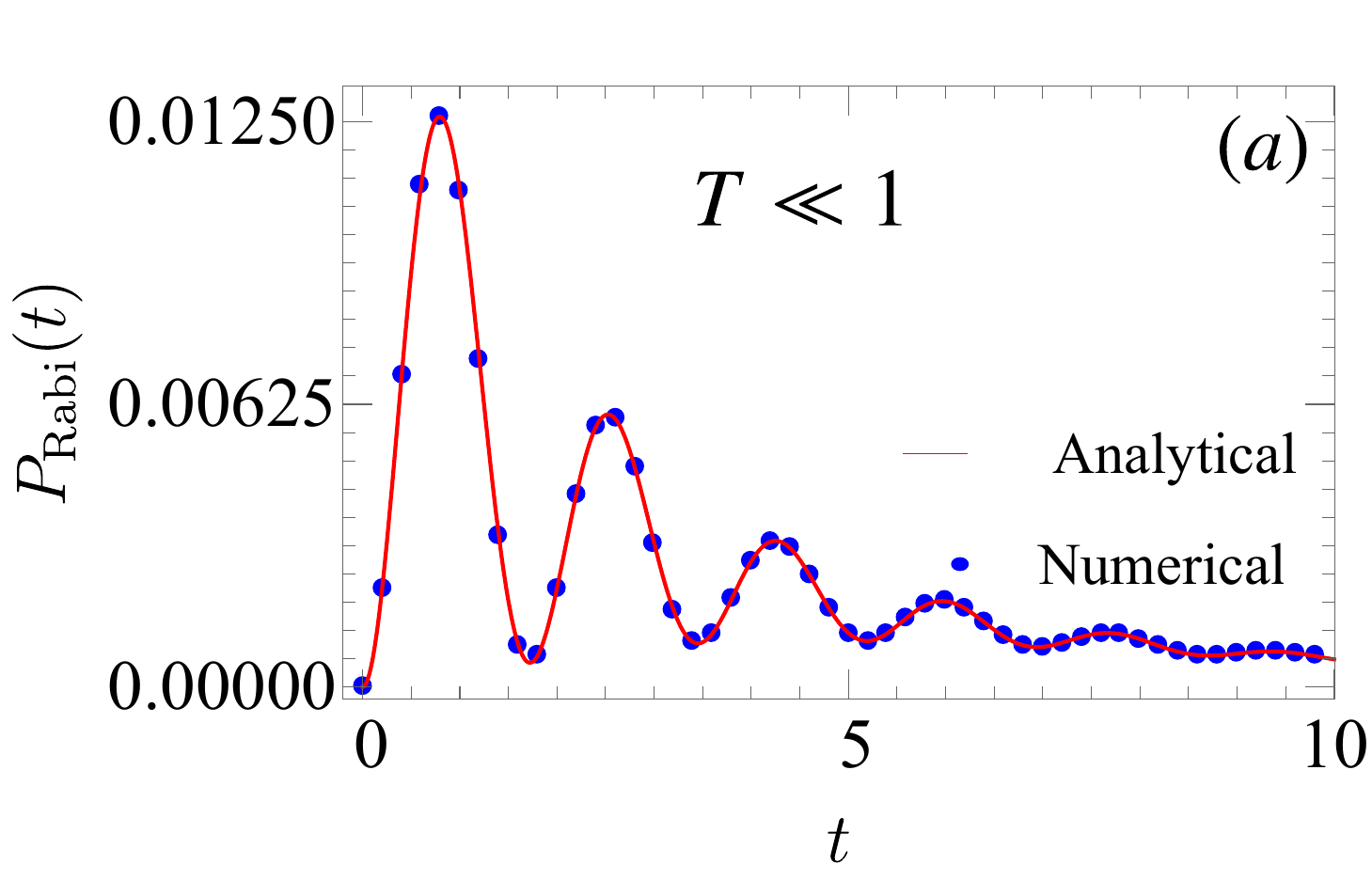}
					\includegraphics[width=7.5cm, height=65mm]{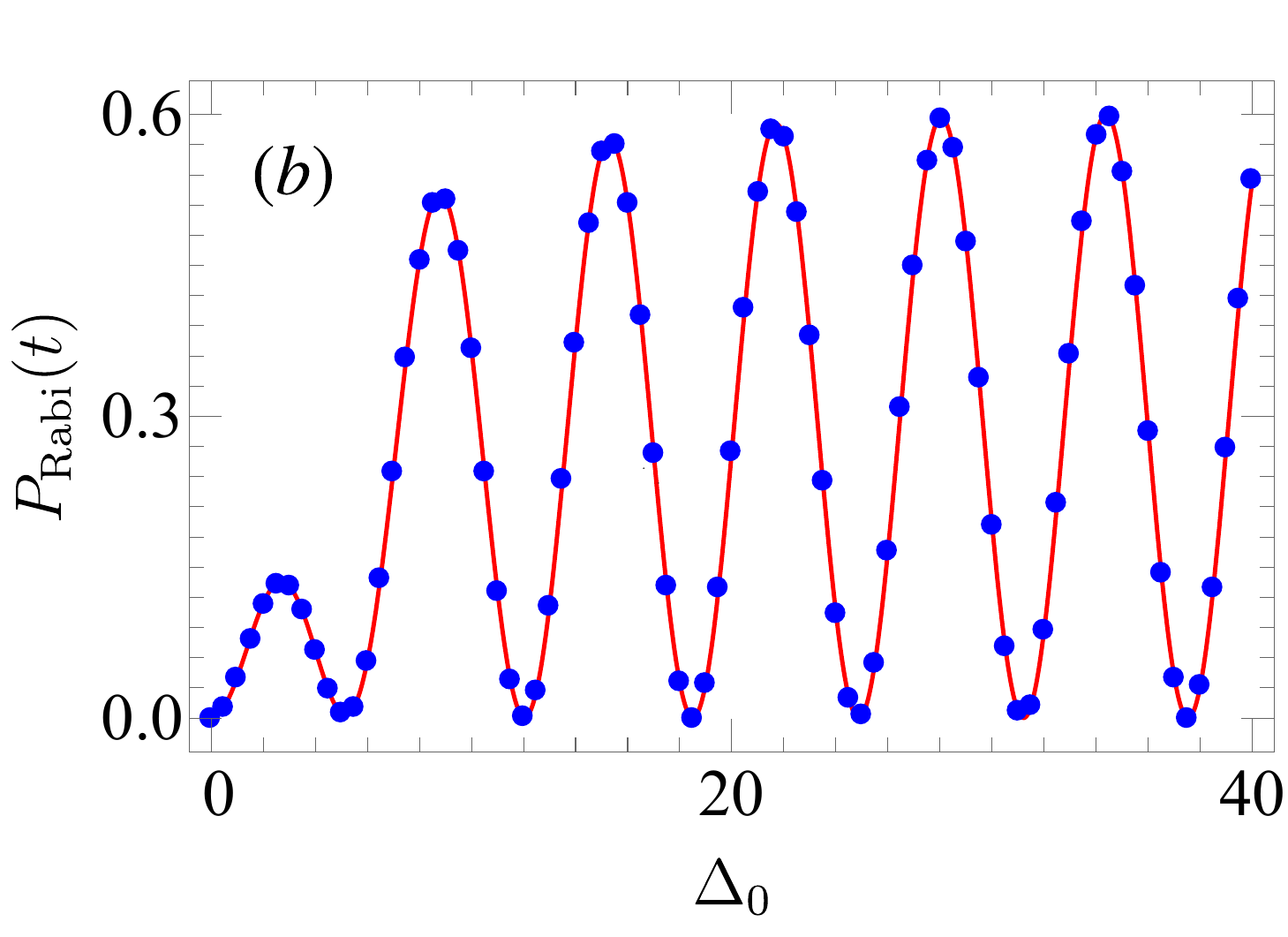}
 \end{center}
\vspace{-0.7cm}
\caption{Correspondence between DK2 and the Rabi model. The excited-state  probability is numerically calculated by solving the time-dependent Schr\"odinger (\ref{equ1}) with the model (\ref{equ2}) in the fast rise limit (small $T$) and by also using equation  (\ref{l3}). For numerical implementation, $T=0.005$,  $\Omega_{0}=0.6$, $\Gamma_{1}=0.8$ and $\Gamma_{2}=0.2$, $\Delta_{0}=0.5$, $D=3$.  The time is in the unit of the second(s) and the initial time is set to $t_{0}=0.0$s. In the panel (b), we have taken $t=1$s } \label{FIG9}
\end{figure}

Following the technique elaborated in Ref.~\onlinecite{Simeon} and assuming that the initial time is $t_{0}=0$,  it can be shown that equation  (\ref{equ28}) for DK2 is tailored by
\begin{eqnarray}\label{l3}
P_{\rm Rabi}(t)=4c^{2}e^{-\bar{\gamma}t}\Big|\frac{\sinh\Big[\frac{1}{2}(\gamma-\alpha-\beta)\ln\frac{2}{\epsilon(t)}\Big]}{\gamma-\alpha-\beta}\Big|^{2},
\end{eqnarray}
where $c$ expresses as in equation  (\ref{equ4.2a}). The parameters $\alpha$, $\beta$ and $\gamma$ are found in equations  (\ref{equ2.7a})-(\ref{equ2.7}). In order to test and confirm the validity of equation  (\ref{l3}), the numerical solution of equation  (\ref{equ1}) with the DK2 model calculated in the fast rise limit is compared with $P_{\rm Rabi}(t)$. The results are depicted on the figure \ref{FIG9}. Both curves are barely discernible confirming that $P_{\rm Rabi}(t)$ is quantitatively and qualitatively correct to characterize a system whose dynamics is encoded into the DK2 model in the fast rise limit. The solution $P_{\rm Rabi}(t)$ in equation  (\ref{l3}) is easy to handle compared to $P_{2}(t, 0)$ in equation  (\ref{equ28}) and both solutions coincide in the fast rise limit. It is worth mentioning from figure \ref{FIG9}(b) that when $\Delta_{0}<|\Gamma_{1}-\Gamma_{2}|/2=0.3$, (real parts of eigen-energies cross while imaginary parts do not, see figure \ref{Comp}) at time $t=10$, in the fast rise limit, populations on the diabatic state $|2\rangle$ are weak. After passing to the regime $\Delta_{0}>|\Gamma_{1}-\Gamma_{2}|/2=0.3$ (imaginary parts of eigen-energies cross while real parts do not)  Rabi oscillations of increasing amplitude occur in the excited-state populations.

\subsection{Slow rise, Landau-Zener model}\label{Sec6.2}

\begin{figure}[!h]
 \begin{center}
					\includegraphics[width=8.5cm, height=65mm]{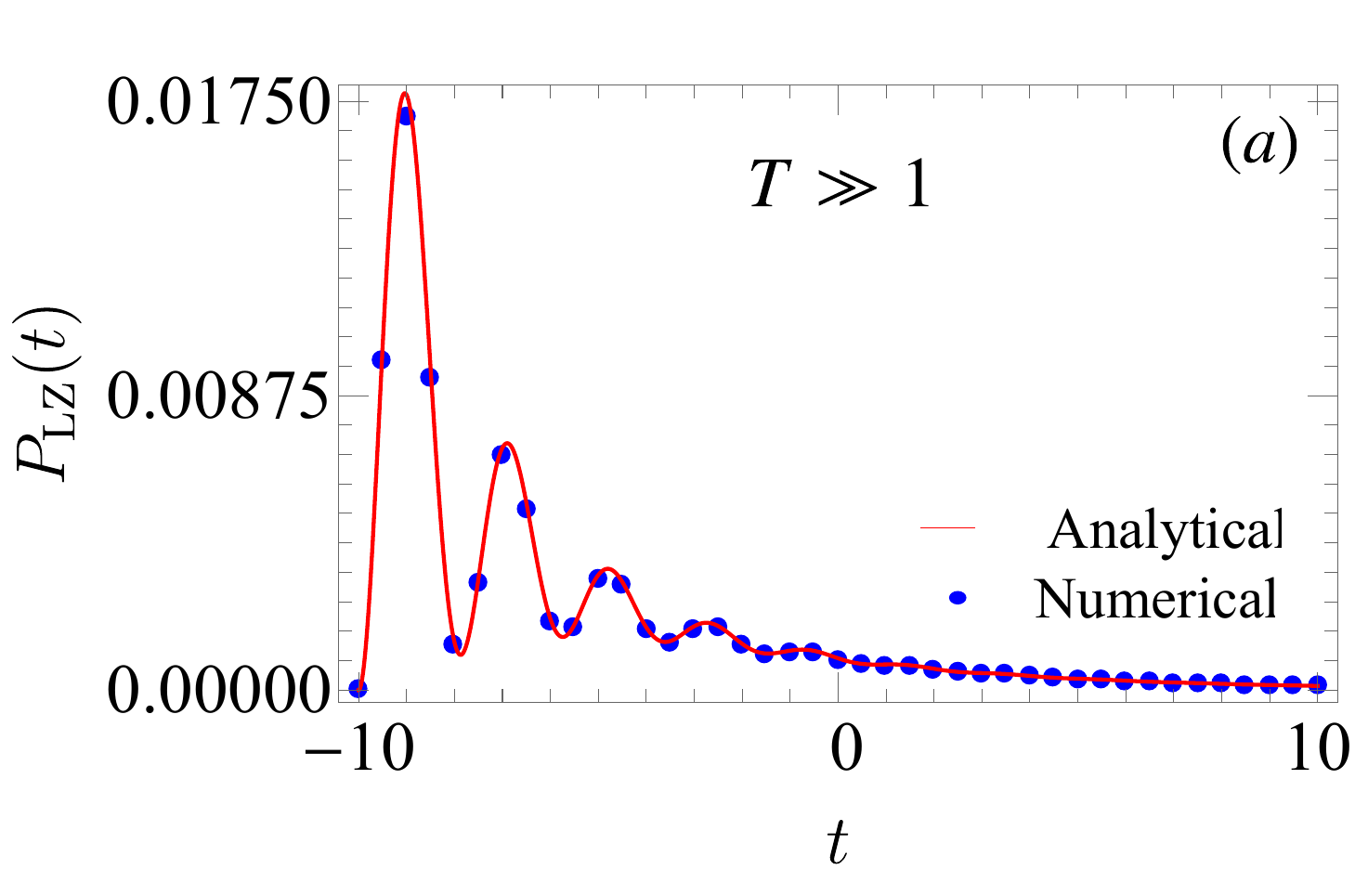}\\\vspace{-0.4cm} 
					\includegraphics[width=8.5cm, height=65mm]{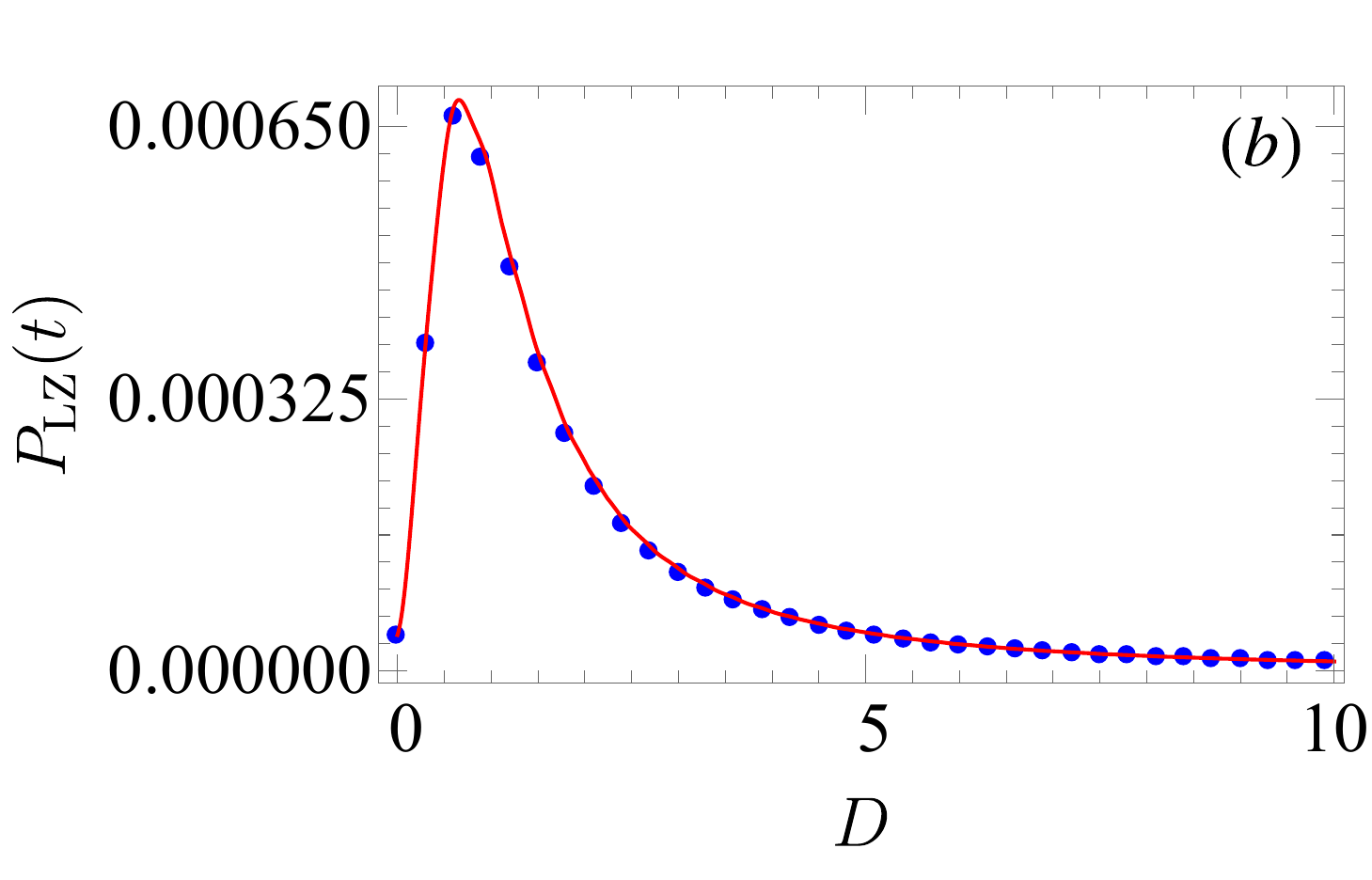}
 \end{center}
\vspace{-0.8cm}
\caption{Correspondence between DK and LZ models. The excited-state probability is firstly calculated by numerically solving the time-dependent Schr\"odinger (\ref{equ1}) with the model (\ref{equ2}) in the slow rise limit (large $T$) and secondly using equation  (\ref{m1}). For numerical implementation, $T=50$,  $\Omega_{0}=0.6$, $\Gamma_{1}=0.8$ and $\Gamma_{2}=0.2$, $\Delta_{0}=0.5$, $D=3$.  The time is in the unit of the second.} \label{FIG10}
\end{figure}

We consider the complementary limit $1/T\to0$ achieved when $T\to\infty$. The time-dependent part of the detuning $\tanh(t/T)\approx t/T$ while the coupling term ${\rm sech}(t/T)\approx 1$ for the DK1 model. In this limit, both DK models become identical and reduce to the Landau-Zener model equation  (\ref{D0}). Still following the method of Ref.\onlinecite{Simeon}, using equation  (\ref{D6}), it can be shown for instance that 
\begin{eqnarray}\label{m1}
\lim_{T\to\infty}z^{\lambda}(1-z)^{\nu}z^{-\lambda}_{0}(1-z_{0})^{-\eta}=e^{-i[\phi(t)-\phi(t_{0})]},
\end{eqnarray}
where the phase $\phi(t)-\phi(t_{0})$ is picked-up by the components of the total wave function during the rising phase of the pulse sweeping from $t_{0}$ to $t$ and
\begin{eqnarray}\label{m2a}
\phi(t)=\frac{vt^{2}}{4}+\frac{\beta_{1}t}{2}.
\end{eqnarray}
However, due to decay,  a direct correspondence between equation  (\ref{equ28}) and an approximated form is not straightforward. One can nevertheless prove that equation  (\ref{equ28}) is conveniently approached by (obtained by directly solving equation  (\ref{equ1}) with the model equation  (\ref{D0}), see Appendix \ref{App3}) 
\begin{eqnarray}\label{m2} 
\nonumber P_{\rm LZ}(y,y_{0})=\Big|(\alpha yy_{0})^{1/2}\frac{\Gamma(\alpha+1)}{\Gamma(\gamma+1)}\Big[\mathcal{J}_{\alpha\gamma}(y,y_{0})-\mathcal{J}_{\alpha\gamma}(y_{0},y)\Big] 
\\\nonumber\times e^{-i[\phi(t)-\phi(t_{0})]-y_{0}}\Big|^{2},\\ 
\end{eqnarray}
where,
\begin{eqnarray}\label{m3}
y\equiv y(t)=\frac{iv}{2}\Big(t+\frac{\beta_{1}-\beta_{2}}{2v}\Big)^{2},
\end{eqnarray}
with
\begin{eqnarray}\label{m4}
v=\frac{\Omega_{0}}{T}, \quad\beta_{1}=D-i\Gamma_{1}, \quad \beta_{2}=-D-i\Gamma_{2},
\end{eqnarray}
and $y_{0}\equiv y(t_{0})$. The function $\mathcal{J}_{\alpha\gamma}(y,y_{0})$ is expressed as a combination of confluent hypergeometric functions 
\begin{eqnarray}\label{m5}
\mathcal{J}_{\alpha\gamma}(y,y_{0})=M(\alpha+1,\gamma+1;y)U(\alpha+1,\gamma+1;y_{0}),\quad
\end{eqnarray}
where $M(...)$ and $U(...)$ are respectively the confluent hypergeometric functions of the first and second kind~\cite{MathBook, Wong}. Likewise,
\begin{eqnarray}\label{m6}
\alpha=\frac{i\Delta^{2}_{0}}{8v}, \quad {\rm and}\quad \gamma=\frac{1}{2}.
\end{eqnarray}
The satisfactory quantitative agreement between numerical and analytical data depicted in the figure \ref{FIG10} confirms the accuracy of our analytical results (\ref{m2}). The relevant probability amplitudes are presented in Appendix \ref{App3}. 

Relevantly, the DK2 model can be linearized at the vicinity of the pseudo-crossing point $t_{cr}=T{\rm arctanh}(-D/\Omega_{0})$ such that near the region $t=t'+t_{cr}$ where the time $t'$ is small, DK2 turns to a LZ model similar to equation  (\ref{D0}). Thus, within the specified region, $\tanh t'/T\approx t'/T$ independently on $T$ and the diabatic energies $\Omega(t')$ and $-\Omega(t')$ linearly cross at $t'=0$ as
\begin{eqnarray}\label{m7}
\Omega(t')\approx vt', \quad {\rm where} \quad v=\frac{\Omega_{0}}{T}\Big(1-\frac{D^{2}}{\Omega^{2}_{0}}\Big).
\end{eqnarray}
Furthermore, these relations establish that, the DK2 and LZ models are not only equivalent and suitable to describe non-adiabatic transitions in the slow rise limit, but also at the vicinity of the pseudo-crossing point. The solution $P_{\rm LZ}(y,y_{0})$ is valid in this case when $y\to y'\equiv y(t')$ and $\beta_{1,2}=-i\Gamma_{1,2}$. 

\subsection{Adiabatic evolutions}\label{Sec7}

In this section, we study the dynamics of the two-level system driven such that it follows one of the eigenstates of the Hamiltonian $\mathbf{H}(t)$ in the absence of decay. This is known as adiabatic evolution.  Such a study is  conveniently performed in the basis of the eigenstates (adiabatic basis) of $\mathbf{H}_{0}(t)$ which is nothing but the Hamiltonian (\ref{equ2}) in the absence of decay.  

In order to construct our adiabatic basis, we consider the eigenstates $|\varphi_{1,2}(t)\rangle$ of the Hamiltonian (\ref{equ2}):
\begin{eqnarray}\label{C1}
|\varphi_{1}(t)\rangle=\cos\vartheta_{\bar{\Gamma}}(t)|1\rangle-\sin\vartheta_{\bar{\Gamma}}(t)|2\rangle,
\end{eqnarray}
\begin{eqnarray}\label{C2}
|\varphi_{2}(t)\rangle=\sin\vartheta_{\bar{\Gamma}}(t)|1\rangle+\cos\vartheta_{\bar{\Gamma}}(t)|2\rangle,
\end{eqnarray}
satisfying $\mathbf{H}(t)|\varphi_{1,2}(t)\rangle=\mathcal{E}_{1,2}(t)|\varphi_{1,2}(t)\rangle$   and its eigenstates $|\varphi_{\pm}(t)\rangle$ in the absence of decay obtained from equations  (\ref{C1}) and (\ref{C2}) by setting $\bar{\Gamma}=0$. They obey the equation $\mathbf{H}_{0}(t)|\varphi_{\pm}(t)\rangle=\mathcal{E}_{\pm}(t)|\varphi_{\pm}(t)\rangle$ where the eigen-energies $\mathcal{E}_{\pm}(t)$ are deduced from equation  (\ref{equ3a}) by putting $\bar{\Gamma}=\bar{\gamma}=0$. 

Because of decay, the eigenstates $|\varphi_{1,2}(t)\rangle$ cannot guarantee population transfer between diabatic states $|1\rangle$ and $|2\rangle$. We  work in an adiabatic basis made up of eigenstates of $\mathbf{H}_{0}(t)$. Thereby, we construct the orthogonal rotation matrix 
\begin{eqnarray}\label{C5}
\mathbf{W}(t)=
\left[
{\begin{array}{*{20}c}
 \cos\vartheta_{0}(t)            &        \sin\vartheta_{0}(t)  \\
        -\sin\vartheta_{0}(t)           &        \cos\vartheta_{0}(t)
\end{array} } \right],
\end{eqnarray}
with the help of which, we rotate the system from diabatic to adiabatic basis through the relation 
\begin{eqnarray}\label{C6}
\mathbf{C}(t)=\mathbf{W}(t)\mathbf{A}(t),
\end{eqnarray}
where $\mathbf{A}(t)=[A_{1}(t,t_{0}), A_{2}(t,t_{0})]^{\mathcal{T}}$ is a two-component vector adiabatic probability amplitude. This scenario suggests that eigenstates $|\varphi_{+}(t)\rangle$ and $|\varphi_{-}(t)\rangle$ of $\mathbf{H}_{0}(t)$ ensure population transfer between two different diabatic states (and not adiabatic states) of $\mathbf{H}(t)$. In this new picture, the Schr\"odinger equation casts the form 
\begin{eqnarray}\label{C7}
 i\frac{d\mathbf{A}(t)}{dt}
=\mathbf{H}_{A}(t)\mathbf{A}(t), 
\end{eqnarray}
where 
\begin{eqnarray}\label{C8}
\nonumber\mathbf{H}_{A}(t)=
\left[
{\begin{array}{*{20}c}
 \lambda_{1}(t)            &        -\frac{i}{2}\Big[\bar{\Gamma}\sin2\vartheta_{0}(t)+\dot{\vartheta}_{0}(t)\Big]  \\
        -\frac{i}{2}\Big[\bar{\Gamma}\sin2\vartheta_{0}(t)-\dot{\vartheta}_{0}(t)\Big]          & \lambda_{2}(t)
\end{array} } \right], \\
\end{eqnarray}
with
\begin{eqnarray}\label{C9}
\dot{\vartheta}_{0}(t)=\frac{\Delta(t)\dot{\Omega}(t)-\Omega(t)\dot{\Delta}(t)}{2[\Delta^{2}(t)+\Omega^{2}(t)]}.
\end{eqnarray}
and
\begin{eqnarray}\label{C10}
\lambda_{1,2}(t)=\frac{1}{2}\Big[\mp\Delta(t)\csc2\vartheta_{0}(t)-i\Gamma_{1,2}\pm2i\bar{\Gamma}\sin^{2}\vartheta_{0}(t)\Big].\qquad
\end{eqnarray}
Note that, $\mathbf{H}_{A}(t)$ is nothing but the Hamiltonian $\mathbf{H}(t)$ in the basis of hybridized eigenstates of $\mathbf{H}_{0}(t)$. 
In adiabatic basis, decay rates of diabatic states $|1\rangle$ and $|2\rangle$ are respectively given by $\Gamma_{1}-2\bar{\Gamma}\sin^{2}\vartheta_{0}(t)$ and $\Gamma_{2}+2\bar{\Gamma}\sin^{2}\vartheta_{0}(t)$.
 
Off diagonal elements of $\mathbf{H}_{A}(t)$ are non-adiabatic couplings. The condition for adiabatic evolution requires that these couplings should be less than energy splitting between levels. For an efficient adiabatic transfer, we assume that adiabatic states $|\varphi_{\pm}(t)\rangle$ of $\mathbf{H}_{0}(t)$ which transport population of diabatic states of the Hamiltonian $\mathbf{H}(t)$ do not perfectly communicate and importantly, are non-degenerate. According to the adiabatic theorem~\cite{Garraway}, this assumption guarantees that, if for instance the system starts at time $t_{0}$ in the diabatic state $|1\rangle$ of the Hamiltonian $\mathbf{H}(t)$ and is slowly transported by the state $|\varphi_{-}(t)\rangle$ of the Hamiltonian $\mathbf{H}_{0}(t)$, when the system reaches the avoided level crossing, there are weak interactions  between $|\varphi_{-}(t)\rangle$ and $|\varphi_{+}(t)\rangle$ such that the most important part of the system ends up in the diabatic state $|2\rangle$ at time $t$ and the remaining part returns to $|1\rangle$ after interactions. In the absence of decay, $|\varphi_{-}(t)\rangle$ and $|\varphi_{+}(t)\rangle$ exactly coincide with $|\varphi_{1}(t)\rangle$ and $|\varphi_{2}(t)\rangle$ respectively. The condition for adiabatic evolution formally reads $\dot{\vartheta}_{0}(t)\ll |\mathcal{E}_{+}(t)-\mathcal{E}_{-}(t)|$ and in the strong adiabatic limit, $\dot{\vartheta}_{0}(t)=0$. Accordingly, we neglect off diagonal elements in $\mathbf{H}_{A}(t)$. Thus, the Schr\"odinger equation 
(\ref{C7}) is readily integrated and returning to the original diabatic basis, one obtains
\begin{eqnarray}\label{C11}
\mathbf{U}_{diab}(t,t_{0})=\mathbf{W}(t)
\left[
{\begin{array}{*{20}c}
 e^{-i\Lambda_{1}(t,t_{0})}  & 0 \\
0  & e^{-i\Lambda_{2}(t,t_{0})}
\end{array} } \right]
\mathbf{W}^{T}(t_{0}),\qquad
\end{eqnarray}
where 
\begin{eqnarray}\label{C12}
\Lambda_{\kappa}(t,t_{0})=\int_{t_{0}}^{t}\lambda_{\kappa}(t')dt'
={\rm Re}\Lambda_{\kappa}(t,t_{0})+i{\rm Im}\Lambda_{\kappa}(t,t_{0}),\qquad
\end{eqnarray}
with $\kappa=1,2$. Here, and from equation  (\ref{C10}), one finds that the real and imaginary parts of the phase $\Lambda_{\kappa}(t,t_{0})$ are given by 
\begin{eqnarray}\label{C13}
\nonumber{\rm Re}\Lambda_{1,2}(t,t_{0})&=&\mp\frac{1}{2} \int_{t_{0}}^{t}dt'\Delta(t')\csc2\vartheta_{0}(t')\\&=&\pm\int_{t_{0}}^{t}dt'\sqrt{\Delta^{2}(t')+\Omega^{2}(t')},
\end{eqnarray}
\begin{eqnarray}\label{C14}
\nonumber{\rm Im}\Lambda_{1,2}(t,t_{0})&=&-\frac{1}{2}\int_{t_{0}}^{t}dt'\Big(\Gamma_{1,2}\mp2\bar{\Gamma}\sin\vartheta_{0}^{2}(t')\Big)\\&=&-\frac{1}{2}\int_{t_{0}}^{t}dt'\Big(\bar{\gamma}\pm\frac{\bar{\Gamma}\Omega(t')}{\sqrt{\Delta^{2}(t')+\Omega^{2}(t')}}\Big).\quad
\end{eqnarray}
Thence, $\mathcal{P}_{1}(t,t_{0})=|\mathbf{U}_{diab}^{11}(t,t_{0})|^{2}$ is the population which returns to $|1\rangle$ after interactions and $\mathcal{P}_{2}(t,t_{0})=|\mathbf{U}_{diab}^{21}(t,t_{0})|^{2}$ represents the population transported to $|2\rangle$ [Here, $\mathbf{U}_{diab}^{\kappa\kappa'}(t,t_{0})$ are matrix elements of $\mathbf{U}_{diab}(t,t_{0})$]. It should be noted that because of decay, $\mathcal{P}_{1}(t,t_{0})+\mathcal{P}_{2}(t,t_{0})\le1$. After calculations, one obtains

\begin{eqnarray}\label{C15}
\nonumber\mathcal{P}_{1}\approx\frac{e^{2{\rm Im}\Lambda_{+}}}{4}\Big[1+\frac{\Omega(t)\Omega(t_{0})}{\omega(t)\omega(t_{0})}\Big]
+\frac{e^{2{\rm Im}\Lambda_{-}}}{4}\Big[\frac{\Omega(t)}{\omega(t)}+\frac{\Omega(t_{0})}{\omega(t_{0})}\Big]\\\nonumber+e^{\Phi_{12}^{\rm decay}}\frac{\Delta(t)\Delta(t_{0})}{2\omega(t)\omega(t_{0})}\cos\Phi_{12}^{\rm dyna},\\
\end{eqnarray}
\begin{eqnarray}\label{C16}
\nonumber\mathcal{P}_{2}\approx\frac{e^{2{\rm Im}\Lambda_{+}}}{4}\Big[1-\frac{\Omega(t)\Omega(t_{0})}{\omega(t)\omega(t_{0})}\Big]
-\frac{e^{2{\rm Im}\Lambda_{-}}}{4}\Big[\frac{\Omega(t)}{\omega(t)}-\frac{\Omega(t_{0})}{\omega(t_{0})}\Big]\\\nonumber-e^{\Phi_{12}^{\rm decay}}\frac{\Delta(t)\Delta(t_{0})}{2\omega(t)\omega(t_{0})}\cos\Phi_{12}^{\rm dyna},\\
\end{eqnarray}
where 
\begin{subeqnarray}\label{C16a}
& e^{2{\rm Im}\Lambda_{\pm}}=e^{2{\rm Im}\Lambda_{1}}\pm e^{2{\rm Im}\Lambda_{2}}, \\\nonumber\\
& \Phi_{12}^{\rm decay}={\rm Im}\Lambda_{1}+{\rm Im}\Lambda_{2},\\\nonumber\\
& \Phi_{12}^{\rm dyna}(t,t_{0})={\rm Re}\Lambda_{1}(t,t_{0})-{\rm Re}\Lambda_{2}(t,t_{0}). 
\end{subeqnarray}
In equations (\ref{C15}) and (\ref{C16}), we have defined   
$
\omega(t)=\sqrt{\Delta^{2}(t)+\Omega^{2}(t)}.
$ 
The above solutions hold for arbitrary pulse and Rabi frequencies. They are fully determined when ${\rm Re}\Lambda_{1,2}(t,t_{0})$ and ${\rm Im}\Lambda_{1,2}(t,t_{0})$ are known. 

\section{Conclusions}\label{Sec8}

We have analytically investigated the dynamics of a decaying two-state system driven by an external pulse whose detuning has a hyperbolic-tangent chirp and a static part. Two special aspects related to possible choices of Rabi frequencies (interactions) are considered. We have separately considered the case when the Rabi frequency is a time-dependent hyperbolic-secant function (first DK model) and the case when it is constant in time (second DK model). Both models are identified as decaying versions of original Demkov-Kunike models. For our analytical treatment, we followed the procedure elaborated in~Refs.[\onlinecite{Avishai2014, Kenmoe2015}]. This allowed us to construct a general theory which copes with both models. We obtained exact analytical solutions. Compared with other procedures developed so far, especially the method in~[\onlinecite{ Garanin}] which considers a semi-classical approach, our results hold for arbitrary decay rates $\Gamma_{1}$ and $\Gamma_{2}$ at arbitrary initial times. 
 
We have analyzed the eigenvalues of the Hamiltonians and establish the condition for the crossing of their real/imaginary parts. The role of the real and imaginary parts of the static detuning is discussed. Thus, when $\Omega_{0}\ge D$ and the half of the difference between imaginary parts  of the static detuning (decay rates $\Gamma_{1}$ and $\Gamma_{2}$) is larger than the energy difference between adiabatic states at time $t_{cr}$ (moment when $\Omega(t_{cr})=0$) in the absence of decays, real parts  of eigenvalues cross while imaginary parts do not. By contrast, when this condition is violated but $\Omega_{0}\ge D$ respected, imaginary parts of eigenvalues cross and real parts do not. Then, the crossing of adiabatic states is attributed to decay of diabatic states. The real parts of the static detuning causes the mixture of population by creating St\"uckelberg oscillations in it (see figures \ref{FIG9} and \ref{FIG10}), while imaginary parts cause the crossing of levels and population destruction. We have also pointed out that the real and imaginary parts of the phase accumulated by the two components of the wave-functions during adiabatic stages are respectively responsible for superposition of states at level-crossing and population destruction. Two complementary limits of the pulse (fast and slow rise) are considered and discussed. Approximated analytical solutions which conveniently fit exact solutions in these limits are obtained and presented.

\section*{Acknowledgments}
The authors thank E. Ngwa, F. Ngoran and A. Kammogne for careful reading of the manuscript and valuable suggestions. One of the authors, M. B. Kenmoe thanks the African Institute for Mathematical Sciences (AIMS) of Ghana, (where the last part of this work was performed) for the hospitality.

\appendix
\section{Mathematical instruments}\label{App1}
The Gauss hypergeometric function $F(\alpha,\beta,\gamma;z)$ is solution to the second-order differential equation~\cite{MathBook, Wong}, 
\begin{eqnarray}\label{01}
 z(1-z)\frac{\partial^{2}F}{\partial z^{2}}+[\gamma-(\alpha+\beta+1)z]\frac{\partial F}{\partial z}-\alpha\beta F=0.\quad
\end{eqnarray}
This equation has two linearly independent solutions $F(\alpha, \beta, \gamma; z)$ and $z^{1-\gamma}F(\beta-\gamma+1, \alpha-\gamma+1, 2-\gamma; z)$ that yield the Wronskian relation 
 \begin{eqnarray}\label{02}
\mathsf{W}=(1-\gamma)z^{-\gamma}(1-z)^{\gamma-\alpha-\beta-1},
\end{eqnarray}
and obeys the derivative properties 
\begin{eqnarray}\label{03}
\frac{d}{dz}F(\alpha,\beta,\gamma;z)=\frac{\alpha\beta}{\gamma}F(\alpha+1,\beta+1,\gamma+1;z),
\end{eqnarray}
and 
\begin{eqnarray}\label{04}
\nonumber\frac{d}{dz}[z^{1-\gamma}F(\alpha-\gamma+1,\beta-\gamma+1,2-\gamma;z)]\\=(1-\gamma)z^{-\gamma}F(\alpha-\gamma+1,\beta-\gamma+1, 1-\gamma;z). 
\end{eqnarray} 
The Gauss hypergeometric functions satisfy the well-known relations 
\begin{eqnarray}\label{05}
\nonumber z^{1-\gamma}F(\alpha-\gamma+1, \beta-\gamma+1, 2-\gamma; z)\\=\mathsf{A}_{1}F(\alpha, \beta, \alpha+\beta+1-\gamma; 1-z)-\mathsf{A}_{2}F(\alpha, \beta, \gamma; z),\quad
\end{eqnarray}
and
\begin{eqnarray}\label{06}
F(\alpha, \beta, \gamma; z)=(1-z)^{\gamma-\alpha-\beta}F(\gamma-\alpha, \gamma-\beta, \gamma; z),\quad
\end{eqnarray}
where
\begin{eqnarray}\label{07}
\mathsf{A}_{1}&=&\frac{\Gamma(\alpha)\Gamma(\beta)}{\Gamma(\alpha+\beta+1-\gamma)\Gamma(\gamma-1)},
\end{eqnarray}
\begin{eqnarray}\label{08}
\mathsf{A}_{2}&=&\frac{\Gamma(\alpha)\Gamma(\beta)\Gamma(1-\gamma)}{\Gamma(\alpha+1-\gamma)\Gamma(\beta+1-\gamma)\Gamma(\gamma-1)}.
\end{eqnarray}
When $z$ asymptotically approaches $1$, then,
\begin{eqnarray}\label{09}
F(\alpha, \beta, \gamma; 1)=\frac{\Gamma(\gamma)\Gamma(\gamma-\alpha-\beta)}{\Gamma(\gamma-\alpha)\Gamma(\gamma-\beta)}.
\end{eqnarray}
 This relation is valid when ${\rm Re}\gamma>{\rm Re}(\alpha+\beta)$.

\section{The functions $\mathsf{U}_{1,2}(z)$ and $\mathsf{V}_{1,2}(z)$}\label{App2}
The aim of this appendix is to present explicit expressions of functions $\mathsf{U}_{1,2}(z)$ and $\mathsf{V}_{1,2}(z)$:
\begin{eqnarray}\label{A1}
\mathsf{U}_{1}(z)=z^{\mu}(1-z)^{\nu}F(\alpha, \beta, \gamma; z),
\end{eqnarray}
\begin{eqnarray}\label{A2}
\nonumber\mathsf{V}_{1}(z)=z^{1+\mu-\gamma}(1-z)^{\nu}F(\alpha-\gamma+1, \beta-\gamma+1, 2-\gamma; z),\\
\end{eqnarray}
\begin{eqnarray}\label{A3}
\nonumber\mathsf{U}_{2}(z)&=&\frac{2i\omega_{0}}{T\Delta}z^{\mu+\theta}(1-z)^{\nu+\theta}\Big[\frac{\alpha\beta}{\gamma} F(\alpha+1, \beta+1, \gamma+1; z)\\ &+&\frac{\mu-(\mu+\nu)z}{z(1-z)}F(\alpha, \beta, \gamma; z)\Big],
\end{eqnarray}
\begin{eqnarray}\label{A4}
\nonumber \mathsf{V}_{2}(z)=\frac{2i\omega_{0}}{T\Delta}z^{\mu-\gamma+\theta}(1-z)^{\nu+\theta}\\\nonumber\qquad\Big[(1-\gamma)F(1+\alpha-\gamma, 1+\beta-\gamma, 1-\gamma; z)\\\nonumber
\hspace{1cm}+\frac{\mu-(\nu+\mu)z}{1-z} F(1+\alpha-\gamma, 1+\beta-\gamma, 2-\gamma; z)\Big]. \\
\end{eqnarray}
As already mentioned in the main text, these functions are structurally identical for both decaying DK models apart from the parameters $\omega_{0}$ and $\theta$ that are $\omega_{0}=1$ and $\theta=1/2$ for DK1, $\omega_{0}=2$ and $\theta=1$ for DK2. The above functions are used to construct matrix elements of the propagator given here by 
\begin{eqnarray}\label{A5}
\nonumber U_{\kappa 1}(z,z_{0})=
\Big[\mathsf{U}_{\kappa}(z)\mathsf{V}_{2}(z_{0})-\mathsf{V}_{\kappa}(z)\mathsf{U}_{2}(z_{0})\Big]\\\times\frac{\exp\Big\{\lambda\ln\frac{z}{z_{0}}+\eta\ln\frac{1-z}{1-z_{0}}\Big\}}{\Lambda(z_{0})},
\end{eqnarray}
\begin{eqnarray}\label{A6}
\nonumber U_{\kappa2}(z,z_{0})=-
\Big[\mathsf{U}_{\kappa}(z)\mathsf{V}_{1}(z_{0})-\mathsf{V}_{\kappa}(z)\mathsf{U}_{1}(z_{0})\Big]\\\times\frac{\exp\Big\{\lambda\ln\frac{z}{z_{0}}+\eta\ln\frac{1-z}{1-z_{0}}\Big\}}{\Lambda(z_{0})},
\end{eqnarray}
$(\kappa=1,2)$ where 
\begin{eqnarray}\label{A7}
\nonumber \Lambda(z_{0})&=&\mathsf{U}_{1}(z_{0})\mathsf{V}_{2}(z_{0})-\mathsf{V}_{1}(z_{0})\mathsf{U}_{2}(z_{0})\\ &=&
\frac{2i\omega_{0}(1-\gamma)}{\Delta T}z^{2\mu-\gamma+\theta}_{0}(1-z_{0})^{2\nu+\gamma-\alpha-\beta-1+\theta}.\quad
\end{eqnarray}
Here, we have used the Wronskian relation (\ref{02}).

\section{The decaying Landau-Zener model, slow rise limit}\label{App3}
In the slow rise limit, the DK models reduce to the LZ model (see discussions in the main text)
\begin{eqnarray}\label{D0}
\mathbf{H}_{\rm LZ}(t)=\frac{1}{2}
\left[ {\begin{array}{*{20}c}
vt+\beta_{1} & \Delta\\
 \Delta & -vt+\beta_{2}
\end{array} } \right],
\end{eqnarray}
where the parameters $v$, $\beta_{1,2}$ are given in equation  (\ref{m4}). Solutions to the  model (\ref{D0}) may be constructed following the procedure elaborated for DK models. As results, probability amplitudes are obtained and written as
\begin{eqnarray}\label{D1}
\hspace{-0.5cm}C_{1,2}=\Big[a_{+}\mathsf{U}_{1,2}(t)+a_{-}\mathsf{V}_{1,2}(t)\Big]\exp\Big[-\frac{i\Omega_{0}t^{2}}{4}-\frac{i\beta_{1} t}{2}\Big].
\end{eqnarray}
The constants $a_{+}$ and $a_{-}$ express as in equations  (\ref{equ17}) and (\ref{equ18}) and the functions $\mathsf{U}_{1,2}(t)$ and $\mathsf{V}_{1,2}(t)$ read
\begin{eqnarray}\label{D2}
\mathsf{U}_{1}(t)&=& M(\alpha,\gamma;y),
\end{eqnarray}
\begin{eqnarray}\label{D3a}
\mathsf{V}_{1}(t)&=& y^{1-\gamma}M(\alpha-\gamma+1,2-\gamma;y),
\end{eqnarray}
\begin{eqnarray}\label{D3b}
\mathsf{U}_{2}(t)&=& -\frac{(\alpha y)^{1/2}}{\gamma}M(\alpha+1,\gamma+1;y),
\end{eqnarray}
\begin{eqnarray}\label{D3c}
\mathsf{V}_{2}(t)&=& -\frac{y^{1/2-\gamma}}{\alpha^{1/2}}(1-\gamma)M(\alpha-\gamma+1,1-\gamma;y),
\end{eqnarray}
$M(...)$ is the confluent hypergeometric function  of the first kind (Kummer's function)~\cite{MathBook, Wong}. It matches the Gauss hypergeometric function as indicted by~\cite{MathBook, Wong}
\begin{eqnarray}\label{D4}
\lim_{|p|\to\infty}F\Big(\alpha,p,\gamma;\frac{y}{p}\Big)=M(\alpha,\gamma;y),
\end{eqnarray}
and relates the confluent hypergeometric functions  of the second kind  through the following relation 
\begin{eqnarray}\label{D5}
\nonumber U(\alpha,\gamma;y)=\frac{\Gamma(1-\gamma)}{\Gamma(\alpha-\gamma+1)}M(\alpha, \gamma; y)\\+\frac{\Gamma(\gamma-1)}{\Gamma(\alpha)}y^{1-\gamma}M(\alpha-\gamma+1, 2-\gamma; y).
\end{eqnarray}
To obtain the expression equation  (\ref{m1}) we have used 
\begin{eqnarray}\label{D6}
\lim_{|p|\to\infty}\Big(1+\frac{y}{p}\Big)^{p}=e^{y}.
\end{eqnarray}
 
\bibliography{Mybib}
\end{document}